\documentclass[reprint,aps,amsmath,amssymb,prf,onecolumn,]{revtex4-2}

\usepackage{physics}
\usepackage{graphicx}
\usepackage{bm}

\usepackage{amsthm} 
\usepackage[T1]{fontenc}
\usepackage{graphicx,tabularx} 
\usepackage{placeins}
\usepackage{natbib} 
\usepackage[dvipsnames]{xcolor}
\usepackage{color}
\usepackage{amsfonts}
\usepackage{siunitx}

\usepackage[utf8]{inputenc}
\usepackage{soul}

\usepackage{geometry}

\graphicspath{{figures/} {figures_Annexes/} {movies/}} 

\usepackage[colorlinks=true, allcolors=blue]{hyperref}

\usepackage{comment} 

\newcommand{\Romain}[1]{\textcolor{teal}{#1}}
\newcommand{\Fred}[1]{\textcolor{brown}{#1}}

\begin{document}
\date{\today}

\title{Effect of wind turbulence on wave generation over a viscous liquid}
\author{R. Mathis$^1$}
\author{S. Cazin$^1$}
\author{J. Methel$^1$}
\author{F. Charru$^1$}
\author{J. Magnaudet$^1$}
\author{F. Moisy$^2$}
\author{M. Rabaud$^2$}
\affiliation{$^1$Univ. Toulouse, Toulouse INP, CNRS, IMFT (Institut de M\'ecanique des Fluides de Toulouse), F-31400 Toulouse, France}
\affiliation{$^2$Universit\'{e} Paris-Saclay, CNRS, FAST, F-91405, Orsay, France}

\begin{abstract}
When wind blows over the surface of a viscous liquid, a clear transition from irregular small-amplitude streamwise-oriented wrinkles to well-defined nearly two-dimensional regular waves is observed at a critical wind velocity. We examine how free-stream turbulence in the air influences the growth of wrinkles and regular waves, as well as the transition between these two regimes. Experiments are carried out in a wind tunnel, in which air is blown over a tank filled with silicone oil whose viscosity is fifty times higher than that of water. The free-stream turbulence is enhanced using upstream grids, achieving relative turbulence intensities up to 8$\%$. Surface deformations are measured using  Free-Surface Synthetic Schlieren with micrometer accuracy. Velocity measurements are performed using hot-wire anemometry above the interface and particle image velocimetry in the liquid. Results reveal two primary effects of grid-enhanced free-stream turbulence: an increase in the wrinkle amplitude, and a reduction in the critical wind speed at the onset of regular waves. Nevertheless, the wrinkle-wave transition still corresponds to an approximately constant friction velocity. Similar to a classical boundary layer over a flat plate, the friction velocity is found to decrease with fetch. From a wave energy balance, we develop a qualitative model explaining why, with the highly viscous liquid considered here, this decrease in the friction velocity results in a non-monotonic variation of the wave amplitude with the fetch. 
\end{abstract}

\maketitle
\section{Introduction}
\label{sec:intro}

The wind–driven generation of surface waves is an apparently simple problem that conceals complex physical mechanisms. Beyond its importance in oceanography and wave forecasting \citep{komen1996dynamics,Janssen_2004,sullivan2010dynamics,ayet2022}, it also plays a central role in industrial processes involving two-phase flows such as oil recovery and coating operation~\citep{Fulgosi2003,Vellingiri2013,Ishimura2025}.  A key aspect of the problem is that the flow on the air side is always turbulent, which makes predictions of the onset velocity and the nature of the initial surface deformations challenging. These early stages indeed depend sensitively on the detailed boundary-layer structure in the air — not only the mean velocity profile but also the fluctuating pressure and shear stress distributions, which are notoriously difficult to control in laboratory experiments.

Since the pioneering works of Phillips  and Miles \citep{Phillips_1957,Miles_1957}, several decades of research have led to
the following two-stage scenario. In a first stage, incoherent wrinkles with very low amplitude are
excited by the turbulent pressure fluctuations present in the boundary layer above the surface (Phillips mechanism~\citep{Phillips_1957}). This results in a linear
increase over time of the energy of the surface displacements, which eventually saturate at a finite amplitude governed by the
liquid viscosity for moderate wind speeds~\citep{Phillips_1957,Gottifredi_1970,Kahma_1988,Zhang_1995,Banner_1998,Caulliez_2008,LiShen2022,LiShen2025}. These wrinkles may be thought of as the superposition of incoherent wakes originating from pressure fluctuations traveling in the turbulent boundary layer in the air~\citep{Perrard2019,Nove_2020}. At low wind speed, typically below $1-$\SI{3}{\meter\per\second} for the air-water system, these structures, referred to as {\it wrinkles} by Paquier et al. \citep{Paquier_2015, Paquier_2016}, reach a statistically stationary state in which the energy injected by pressure fluctuations in the surface deformations is balanced by that dissipated in the liquid. This statistical equilibrium corresponds to the asymptotic regime described by Phillips inviscid resonant theory \citep{Phillips_1957}.

In the second stage, for sufficient wind speeds, the coupled air-water shear flow becomes unstable, leading to a subsequent exponential increase of the energy of the surface displacements
(Miles mechanism~\citep{Miles_1957}). 
Linear stability theory can be used to predict the critical wind speed corresponding to this transition~\citep{Miles_1957, Miles1962generation,Gastel1985phase,chaubet2024effect}, as confirmed in laboratory experiments~\citep{Kawai1979generation, veron2001experiments,PEIRSON:2008,Geva2022,Kumar2024}.
The onset is expected to be sensitive to several factors, in particular the liquid viscosity and the free-stream turbulence.
The effect of the liquid viscosity  has been investigated by Paquier {\it et al.}~\citep{Paquier_2016}, who showed that increasing the viscosity increases the critical wind speed.
In contrast, the effect of the free-stream turbulence on wrinkles and critical wind speed remains largely unknown.

Despite their very small amplitude,  wrinkles are the base state from which regular waves grow as the wind speed is increased. Therefore, one may expect the transition to regular waves to depend on any parameter that may affect the wrinkles, in particular the structure of the turbulent boundary layer in the air.
The relation between the statistical properties of the turbulent pressure fluctuations and that of the wrinkles was analyzed in Perrard {\it et al.}~\citep{Perrard2019}. The characteristic size of the latter, $\Lambda$, is governed by the largest spatial scales of the pressure fluctuations field, which itself is controlled by the thickness of the boundary layer, $\delta$, with no significant effect of the liquid viscosity, $\nu_\ell$. The characteristic amplitude of the wrinkles, $\zeta_{\mathrm{rms}} = \langle \zeta^2 \rangle^{1/2}$, with $\zeta({\bf x},t)$ the surface displacement field, results from a balance between the work performed by the wind stress at the surface and the viscous dissipation in the liquid. This balance yields \citep{Perrard2019} 
\begin{equation}
\frac{\zeta_{\mathrm{rms}}}{\delta}\propto\frac{\rho_a}{\rho_\ell}\frac{u^{*3/2}}{(g\nu_\ell)^{1/2}}\,,
\label{eq.perrard}
\end{equation}
with $u^*$, $\rho_a$ and $\rho_\ell$ being the friction velocity in the air and the air and liquid densities, respectively, $g$ denoting gravity.
Two important assumptions are involved in the derivation of \eqref{eq.perrard}. First, the characteristic size of the turbulent pressure fluctuations is governed by the boundary layer thickness $\delta$. Second, the amplitude of pressure fluctuations is governed by the mean shear stress $\rho_a u^{*2}$. These are natural assumptions in standard boundary layers developing over a flat plate with zero or negligible free-stream turbulence. In contrast, the relationship between the statistics of the near-surface pressure fluctuations (size and amplitude) and the shear stress may be more complex in boundary layers with significant free-stream turbulence. The question therefore arises of whether the scaling of \eqref{eq.perrard} still holds in such non-standard boundary layers. 

The purpose of this study is to explore how free-stream turbulence influences the development of wrinkles and the critical wind speed for the onset of wave amplification.
In particular, we examine whether this critical wind speed is modified by changes in the properties of the incoming flow.
If yes, this would imply that wind velocity alone does not control wave onset for a liquid with given physical properties.
This would provide a natural explanation for the large scatter in reported critical wind velocities in the literature for the classical air–water configuration.

To address these questions, we have developed a dedicated wind-tunnel experiment in which air is blown through a grid before encountering the surface of a viscous liquid (silicone oil with kinematic viscosity $\nu_\ell$ = \SI{50e-6}{\meter\squared\per\second}). Two grids with different mesh sizes are used to vary the characteristics of the free-stream turbulence. Velocity measurements are carried out using hot-wire anemometry above the interface and particle image velocimetry within the liquid, enabling a direct determination of the friction velocity $u^*$. Surface deformations are measured using Free-Surface Synthetic Schlieren with micrometer accuracy~\citep{Moisy09}.

The structure of the paper is as follows. 
Section~\ref{sec:setup} presents the experimental set-up, the grid characteristics and the experimental techniques. Section~\ref{sec:flows} provides an overview of the mean velocity profile and turbulence intensity in the air boundary layer, together with the velocity profile in the liquid used to determine the friction velocity at the interface. 
Section \ref{sec:FSTeffect} analyses the statistics of the surface deformations for different wind speeds and free-stream turbulence conditions. Specific attention is paid to the spatial evolution of the wave amplitude with the distance from the grid. The observed non-monotonic evolution is rationalized  thanks to a model based on a wave energy balance taking into account the decrease of the friction velocity in the air with the distance from the grid.

\section{Experimental set-up and measurement techniques}
\label{sec:setup}

\begin{figure}[t]
\centering
\includegraphics[width=\textwidth]{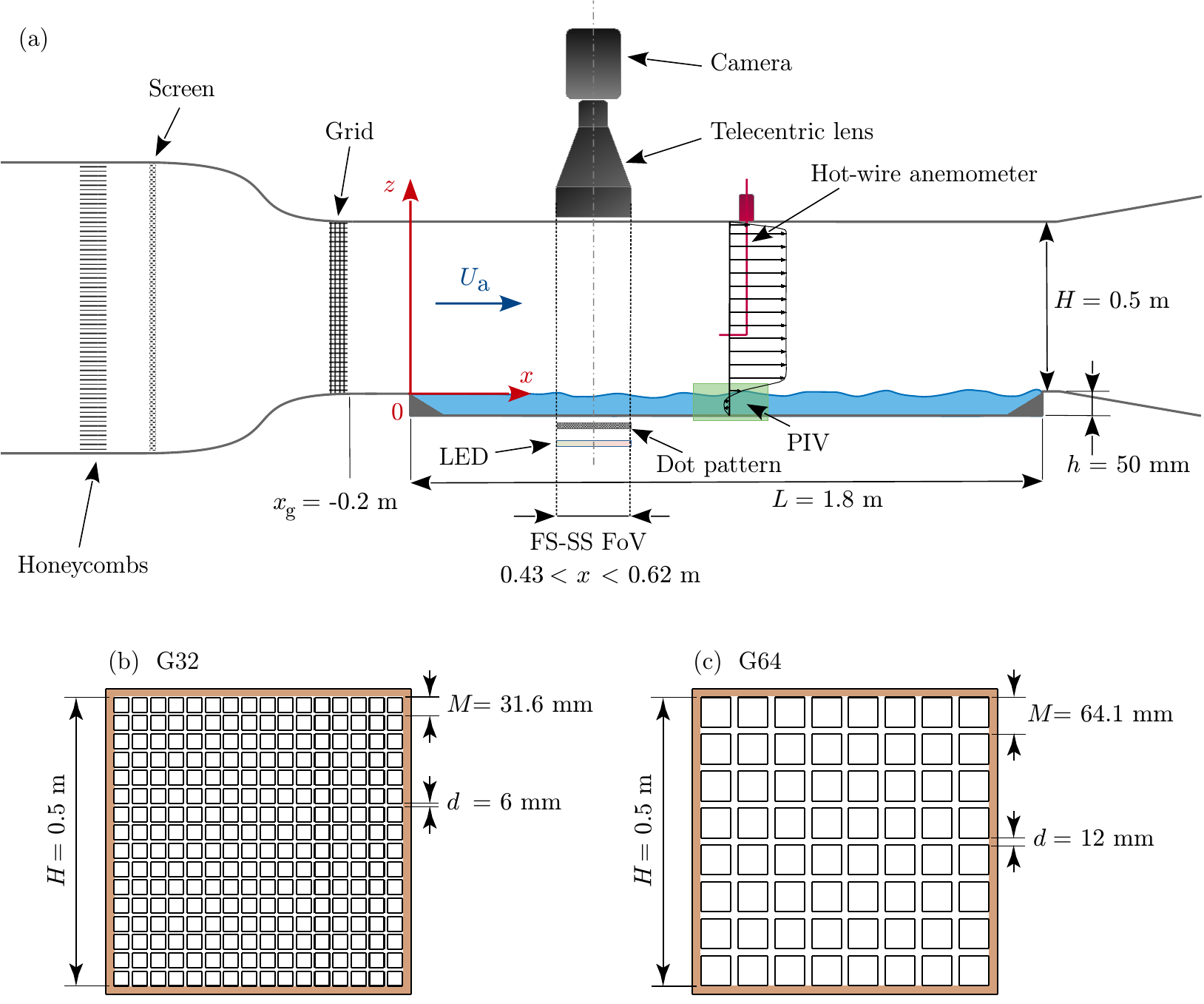}
\caption{Experimental setup. (a) Wind tunnel, with the 50\,mm-deep tank filled with silicone oil and the three measurement techniques. (b) and (c) Grids G32 and G64, respectively. The edges of the grids (in brown) are recessed into the channel walls.}
\label{fig:setup}
\end{figure}

The experimental setup is sketched in Fig.~\ref{fig:setup}a. Experiments are carried out in the closed-loop wind tunnel S2 at IMFT, whose test section is 1.8 m long with a  0.5 m $\times$ 0.5 m cross section. The general characteristics of the wind tunnel are reported in Ref.~\citep{bourguet23_JFS}, and only modifications introduced for the present investigation are described below. The floor of the test section is replaced with a rectangular liquid tank with a depth $h =$ \SI{50}{\milli\meter} spanning the entire width of the tunnel.

The tank is filled with silicone oil with density $\rho_\ell=$ \SI{970}{\kilo\gram\per\cubic\meter} and kinematic viscosity $\nu_\ell=$\SI{50e-6}{\meter\squared\per\second} at \SI{25}{\degreeCelsius}. This high viscosity ensures a strong attenuation of the wave field, preventing wave reflections at the downstream end of the tank and on the side walls.
The side walls of both the test section and the liquid tank are made of glass to provide optical access. In the following, $x$, $y$ and $z$ denote the streamwise, spanwise and vertical directions, with origins located at the leading edge of the tank, at the middle span of the test section, and at the liquid/air interface, respectively. By definition, $x$ also corresponds to the local fetch, i.e., the distance over which the wind has acted on the liquid surface since the onset of air–liquid interaction. To ensure that the flow in the liquid is fully developed, data acquisition starts 20 mins after the wind starts blowing.

Three distinct inflow conditions are investigated: one without a grid, referred to as NG (no-grid), and  two with grids, with mesh sizes $M = 31.6$\,mm and $64.1$\,mm, hereinafter referred to as G32 and G64, respectively. The grids are located 0.2 m downstream of the exit of the contraction, at position $x_g=-0.2$ m upstream of the leading edge of the tank. In what follows, $X=x-x_g$ denotes the streamwise distance to the grid location.
Both grids are made of 8 mm-thick alloy co-planar rectangular bars arranged in a square mesh array (see Figs.~\ref{fig:setup}b and ~\ref{fig:setup}c). 
Grids G32 and G64 are based on bars with width $d=6\,$mm and 12 mm, respectively. They have identical porosity ratios $\beta=(1-d/M)^2=0.66$. Each grid is flush-mounted, with the edge recessed into the walls of the wind tunnel. 

The turbulent boundary layer in the air is characterized by hot-wire anemometry for the three inflow conditions. 
A miniature single-wire probe is used, consisting of a \SI{5}{\micro\meter}-diameter and 1.25 mm-long tungsten wire (DANTEC-55P11), operated by an A. A. Labs AN-1003 anemometer. 
The sampling frequency and duration are set to 20 kHz and 240 s, respectively. The wire sensor is mounted on a dual-axis traverse system that allows a continuous probe motion either in the longitudinal $(x,z)$ plane or in the transverse $(y,z)$ plane. Hot-wire calibration is performed in-situ, in the absence of a grid.

The flow in the liquid tank is measured using Particle Image Velocity (PIV) in the $(x, z)$ vertical mid-plane located at the spanwise position $y=0$ (Fig.~\ref{fig:setup}). 
The PIV setup is composed of a Nd-YAG laser Quantel Ultra 2 $\times$ 200 mJ and a 16-bit LaVision Imager sCMOS camera with a resolution of  2560  $\times$ 2160 pixels, equipped with a 50 mm lens.
In most cases, a $10\,$minutes-long set of 600 images is recorded at a sampling rate of 1 Hz to ensure statistical convergence.
The time delay between successive image pairs is adjusted according to the wind speed and interface roughness, ranging from $200\,$ms for a low wind speed and nearly flat interface to $10\,$ms for high wind speeds and strongly deformed interface.
At low wind speed (1 and \SI{2}{\meter\per\second}), the number of images in each set is increased to 2000 due to the slow convergence rate of statistics in such configurations. Similarly, in the wave regime, the number of images in each set is increased to 1000 in order to obtain converged statistics of  the interface displacement.
Velocity fields are measured in four 324\,mm-long windows with a 44\,mm overlap, covering streamwise positions ranging from $x=100$\,mm to $x=1164$\,mm,
with a spatial resolution of the order of $0.5$ mm, both along the $x$ and $z$ directions. This field of view includes the region where the surface displacements are measured. 
PIV calculations are performed using the DaVis software, using a multi-pass adaptive interrogation scheme with window sizes decreasing from $64\times64$ to $16\times16$ pixels and a 75\% overlap to compute correlation functions.


Surface displacements are measured using the Free Surface Synthetic Schlieren (FS-SS) technique \citep{Moisy09}.
A 25\,cm $\times$ 25\,cm random dot pattern is placed beneath the liquid tank and illuminated by a LED panel. The image of the pattern, refracted through the deformed interface, is recorded using a 2016$\times$2016 pixel, 12-bit pco.dimax cs4 CMOS camera located above the surface (Fig.~\ref{fig:setup}). The camera is equipped with a telecentric lens with a 260 mm entrance diameter (Opto Engineering TC4M-172), offering a wide field of view and a large working distance (600 mm) enabling imaging without parallax distortion. The distance between the pattern and the free surface is adjusted based on the typical wave amplitude to be measured, so as to maximise the measurement accuracy while avoiding the formation of caustics~\citep{Moisy09}. 

A digital image correlation algorithm (DaVis software with the same multi-pass adaptive interrogation scheme as in the PIV above)  is used to determine the apparent displacement field, $\delta {\bf r}$, of the dot pattern relative to a reference image obtained in the absence of wind. 
The free-surface height, $\zeta(x,y)$, is then computed by inverting the linearized refraction problem, ${\boldsymbol \nabla} \zeta = \delta {\bf r} / H^*$, with $H^*$ being the effective distance between the pattern and the surface corrected for the refraction indices of the liquid layer and glass bottom wall. The horizontal field over which $\delta {\bf r}$ is recorded is 190\,mm long and 150\,mm wide, covering a range of fetches from $430\,\text{mm}$ to $620\,\text{mm}$, 
centered at $y=0$. The horizontal resolution is 0.86\,mm while the vertical resolution is approximately \SI{0.1}{\micro\meter}. 
Data are recorded at a rate of 2\,Hz during 500\,s to ensure statistical convergence of the root mean square amplitude, $\zeta_{rms}$.

\section{Flow characterization}
\label{sec:flows}

\subsection{Wind velocity profile and free-stream turbulence}
\label{subsec:wind}

\begin{figure}[t]
\centering
\includegraphics[width=0.95\textwidth]{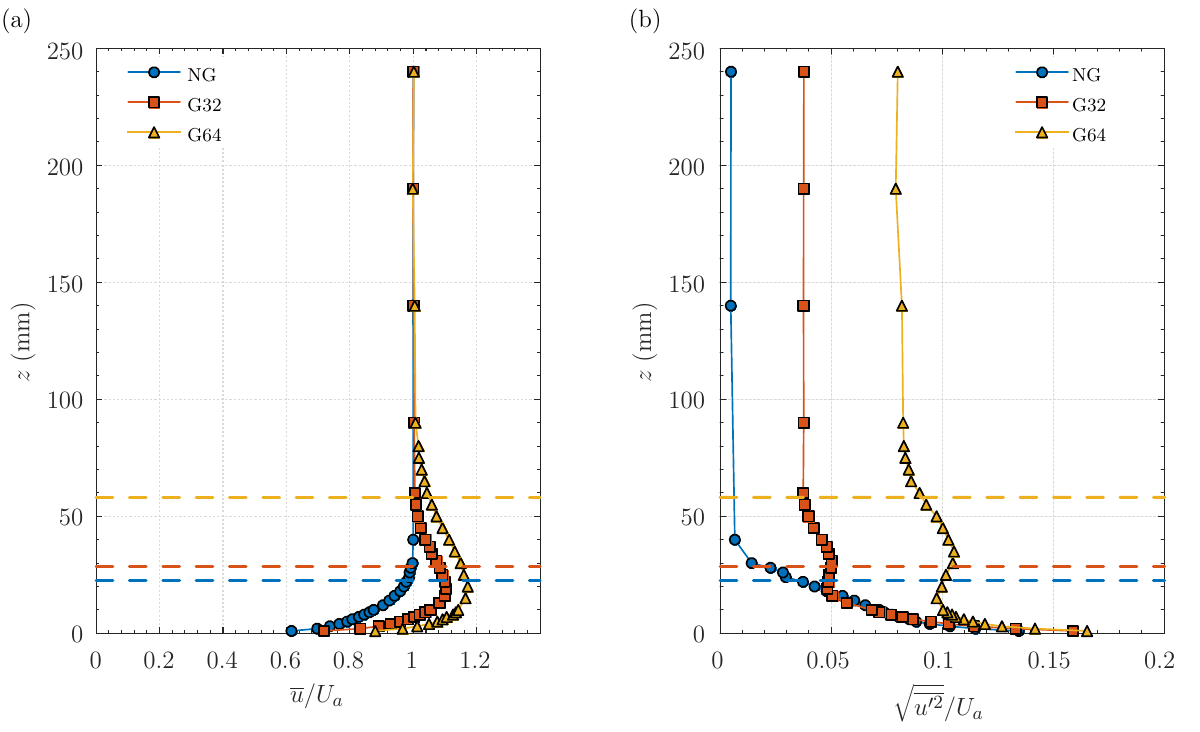}
\caption{Normalized mean velocity profiles (a) and turbulence intensity  profiles (b) in the air at $y=0\,$mm and $x=430\,$mm, \textit{i.e.}, $X=x-x_g = 630\,$mm downstream of the grid location, for the free-stream velocity $U_a=$ \SI{3.2}{\meter\per\second}.
The horizontal dashed lines correspond to the boundary layer thickness for the no-grid case (NG, blue line), and to the location of the center of the first horizontal rod for grids G32 and G64 (red and yellow dashed lines, respectively).}
\label{fig:HWprofiles}
\end{figure}

We first characterize the structure of the turbulent boundary layer in the air and examine its dependence on the inlet conditions.
Figure~\ref{fig:HWprofiles}(a) presents the mean velocity profile $\overline{u}(z)$ for the three inlet conditions at the same free-stream velocity $U_a =$ \SI{3.2}{\meter\per\second}. Profiles are measured at a fetch $x = 430\,\mathrm{mm}$ corresponding to a distance $X = x-x_g = 630\,\mathrm{mm}$ from the grid, i.e., 20 and 10 mesh sizes downstream of grids G32 and G64, respectively.

The no-grid profile exhibits the expected classical behavior, with a boundary layer thickness $\delta \simeq 22\,$mm (here $\delta$ is defined as the $99\%$ boundary layer thickness). In contrast, the presence of a grid markedly alters the velocity profiles, producing a $10-20\%$ overshoot of $\overline{u}(z)$ at a height $z_{\max} \simeq 20$\,mm above the wall, with only a weak dependence on the mesh size. Far from the wall, the velocity is uniform, without any sign of a disturbance resulting from the presence of the grid. This indicates that the grid-induced wake and jet modulation are smoothed out at a downstream distance $X/M \simeq 10-20$, in line with available observations \citep{cardesa12_EiF}. Additional measurements at other spanwise locations confirm the absence of grid-induced modulations in the velocity profile.

The maximum in the velocity profile is a classical feature of turbulent boundary layers in the presence of grids~\citep{owen57_JFM, lau68_JFM, mehta85_AIAAj, irps2016interaction}. It originates from the localized  pressure loss induced by the grid in the free stream, which accelerates the flow in the boundary layer~\citep{owen57_JFM}. The overspeed ratio $U_{max} / U_a$, with $U_{max}=\overline{u}(z_{max})$, is almost independent of the free-stream velocity and exhibits a weak decrease with the streamwise position, not exceeding 3\% over the field of view with both grids. The vertical location of the maximum, $z_{max}$, decreases with $U_a$ by approximately $-8\%$ per \SI{}{\meter\per\second} and  increases with the downstream position by $12\%$ per $100\,$mm. 

Vertical profiles of the streamwise turbulent fluctuation $T_u=\sqrt{\overline{u^{\prime 2}}}/U_a$ are shown in Fig. \ref{fig:HWprofiles}b. In the absence of a grid, these profiles exhibit the classical large inner peak very close to the wall, where $T_u \simeq 15\%$, while $T_u$ decays to $\simeq 0.6\%$ outside the boundary layer. In the presence of a grid, $T_u$ is still uniform outside the boundary layer but reaches much larger free-stream values, $T_u \simeq 4\%$ and $8\%$ with grids G32 and G64, respectively. Close to the wall, $T_u$ keeps values similar to those observed in the no-grid case, with $T_u \simeq 15\%$, indicating that the near-wall fluctuations are dominated by the intrinsic dynamics of the turbulent boundary layer, rather than by the  diffusion from the free-stream turbulence. A local minimum of $T_u$ may be noticed at a vertical position close to $z_{max}$, together with a weak secondary peak at $z \approx 30\,$mm. These are classical features of turbulent shear flows, associated with the strong correlation between the turbulent energy production and the local mean shear. Additional measurements at larger fetch (not shown) exhibit similar trends, with a slow longitudinal decrease of the secondary peak of $\overline{u'^2}$, accompanied by a slight increase of its wall-normal position.

\begin{figure}[t]
\centering
\includegraphics[width=0.98\textwidth]{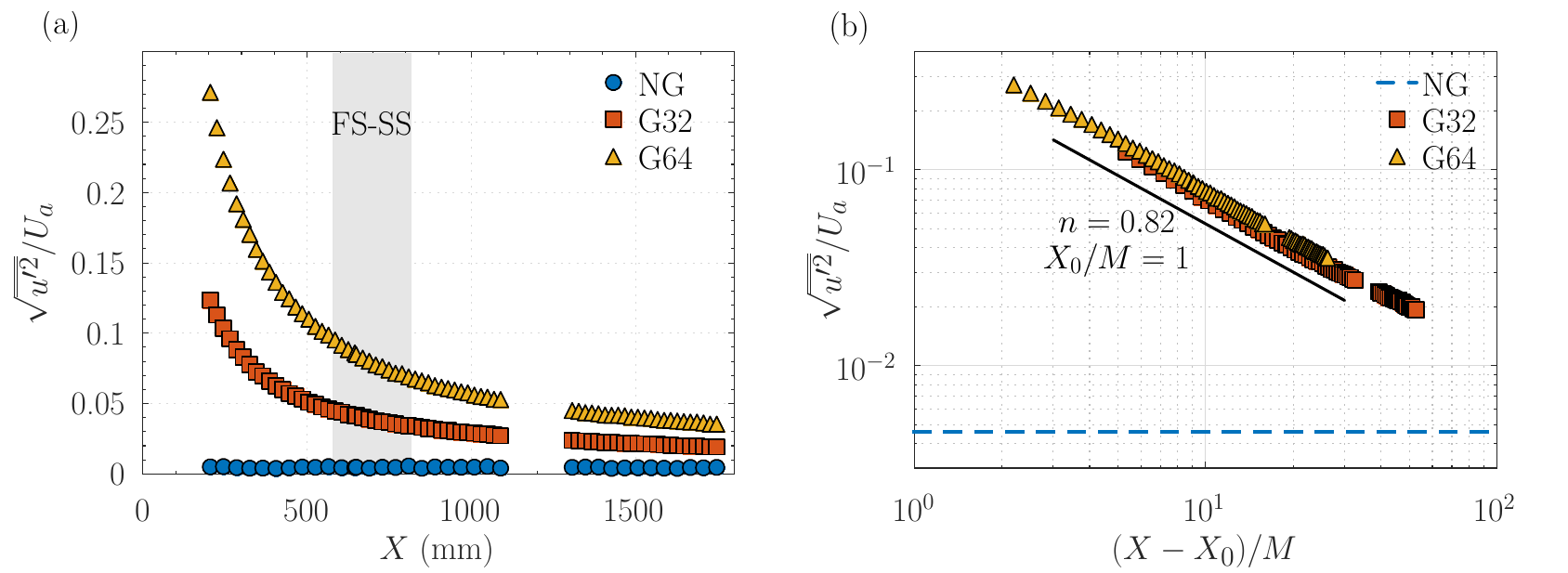}
\caption{Streamwise variation of the free-stream turbulence intensity for $U_a=$ \SI{3.2}{\meter\per\second} (hot wire measurements). (a) Linear plot versus the distance to the grid. The light grey area  indicates the region where the FS-SS measurements of the free-surface displacements are performed. (b) log-log plot against $(X-X_0)/M$, with $X_0/M=1$. The corresponding Reynolds numbers are $Re_M= 7500$ for G32 and 14500 for G64. The dashed blue line represents the uniform free-stream turbulence level (0.6\%) in the NG case. }
\label{fig:Tu}
\end{figure}

Figure~\ref{fig:Tu}(a) shows the variation of the free-stream turbulence intensity $T_u$ with the distance from the grid at $U_a =$ \SI{3.2}{\meter\per\second}. Measurements are performed upstream of, within, and downstream from the FS–SS window where free-surface displacements are determined (grey area in the figure). Without a grid, the turbulence level remains constant, with $T_u \simeq 0.6\%$. It reaches much higher values and decays with the fetch in the presence of a grid, with typical levels $T_u \simeq 3- 4\%$ (G32) and $7- 9\%$ (G64) in the FS-SS window.
The observed decay of the turbulence intensity can be compared with the classical power-law behavior of grid-generated turbulence,
\begin{equation}
T_u= \frac{\sqrt{\overline{u'^2}}}{U_a} = a \left(\frac{X-X_0}{M} \right)^{-n},
\end{equation}
with $X_0$ a virtual origin, $n$ the decay exponent and $a$ a pre-factor. Present data, shown in log–log scale in Fig.~\ref{fig:Tu}b, are consistent with this power-law decay. A best fit with $X_0$ and $n$ as free parameters yields $X_0/M \simeq 1$ and $n \simeq 0.82 \pm 0.02$.
Values of $n$ reported in the literature are usually such that $0.55<n<0.7$, so that present data lie somewhat beyond the upper limit of the usual range~\citep{mohamed90_JFM, lavoie07_JFM, zhao23_EJMBF}.
This may be due in part to the short downstream distances used in present experiments, $X/M \simeq 2-50$, while much larger distances ($X/M > 40$) are usually considered in the literature. These unusually small distances are necessary here to ensure high turbulence intensities, hence a significant effect of the free-stream turbulence on the wave generation process.


\subsection{Influence of the inlet condition on the friction velocity}
\label{subsec:liquid}

\begin{figure}[t] 
\centering
\includegraphics[width=0.8\textwidth]{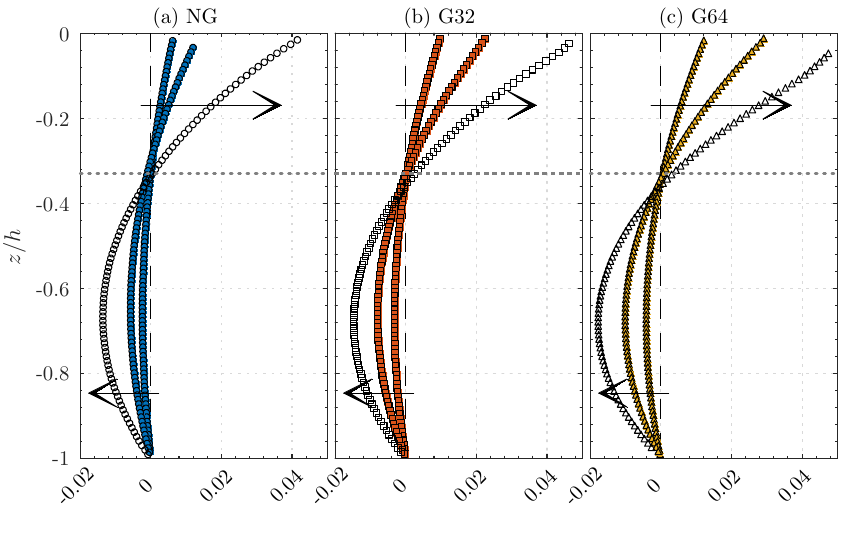}
\caption{
Mean velocity profile in the liquid determined by PIV at fetch $x=500\,$mm, in the absence (a) or in the presence (b,c) of a grid. Arrows indicate increasing wind speeds $U_a$. Filled symbols correspond to wind speeds below the onset of waves ($U_a=3.2$ and 5.1~\SI{}{\meter\per\second} for the three inlet conditions) while open symbols refer to wind speeds beyond the onset of waves ($U_a=8.3$, 6.1 and 5.6 \SI{}{\meter\per\second} for NG, G32 and G64, respectively).}
\label{fig:PIVumean}
\end{figure}

A key quantity in the surface wave generation process is the wind shear stress $\tau = \mu_a \partial \overline{u}/\partial z$ at the air-liquid interface, usually quantified through the friction velocity $u^* = \sqrt{\tau / \rho_a}$. 
Since the minimum distance above the interface reached in hot-wire measurements is not sufficient to extrapolate the velocity gradient down to the interface, $\tau$ is usually obtained indirectly. In a classical boundary layer (with negligible free-stream turbulence), $u^*$ is inferred by fitting the mean velocity profile with the 
logarithmic law. However, this law no longer applies in non-standard boundary layers with a strong free-stream turbulence. Nevertheless, we can deduce $u^*$ from the shear-induced flow in the liquid, taking advantage of the continuity of shear stresses at the interface \citep{Paquier_2015}. Because of the large liquid viscosity, the flow induced in the liquid by the wind stress is laminar in present experiments (the maximum Reynolds number is $U_s h / \nu_\ell \simeq 40$, with $U_s$ being the drift velocity at the surface and $h$ the liquid depth). Assuming a purely two-dimensional flow, the zero net flow-rate condition implies a parabolic Couette-Poiseuille profile~\citep{spurk08_BOOK}
\begin{equation}
\label{eq:parab}
   u_\ell(x,z) = U_s(x)\left(1+\frac{z}{h}\right)\left(1+3\frac{z}{h}\right),
\end{equation}
with $-h < z < 0$. Continuity of shear stresses at the air-liquid interface implies
\begin{equation}
\label{eq:utau}
    \rho_a u^{*2} = \rho_\ell \nu_\ell \frac{\partial {u_\ell}}{\partial z} \big|_{z=0} \,.
\end{equation}

Velocity profiles determined in the liquid by PIV in the vertical midplane $y=0$ are shown in Fig. \ref{fig:PIVumean} for various wind speeds and for the different inflow conditions.
At low wind speeds (filled symbols), the interface remains almost flat and the flow is stationary. At higher wind speeds (open symbols), the surface becomes wavy, and the instantaneous velocity profiles exhibit a time-dependent oscillatory component. In this regime, a time average is applied to extract the mean velocity profile. In both cases, the resulting profiles are accurately captured by the parabolic law \eqref{eq:parab}. More specifically, profiles obtained in the absence of a grid exhibit a zero-crossing at $|z|/h \in [0.30, 0.33]$, close to the expected value $1/3$, suggesting that the flow is almost two-dimensional and thus uniform in the spanwise direction. Departures from the theoretical zero-crossing position are more pronounced in the presence of grids, where $|z|/h \in [0.34, 0.36]$, and slightly more pronounced with G64 (Fig. \ref{fig:PIVumean}c) than with G32 (Fig. \ref{fig:PIVumean}b). 
This suggests the presence of a positive flow rate in the midplane $y=0$, implying a weak return flow close to the sidewalls. However, the resulting spanwise modulation of the surface velocity remains small and is unlikely to affect the wave generation process.

\begin{figure}[t]
\centering
\includegraphics[width=0.8\textwidth]{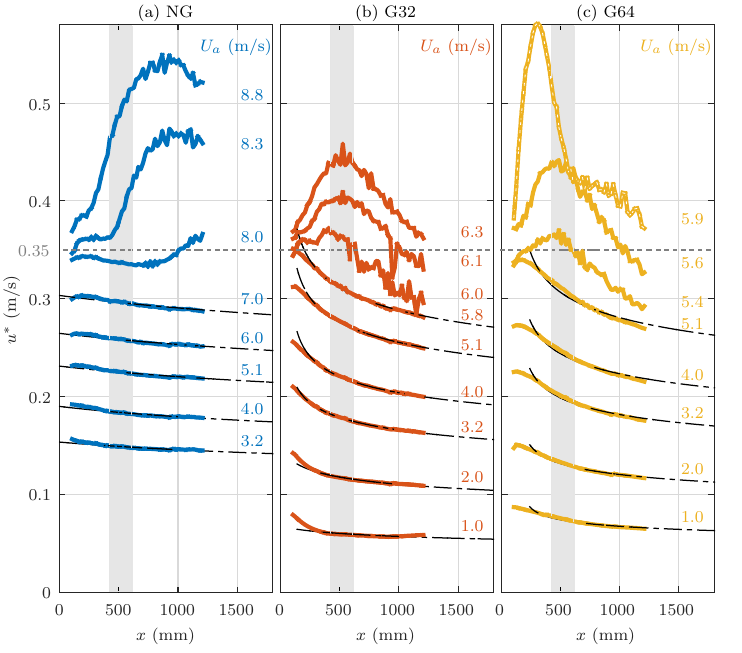}
\caption{
Longitudinal variation of the friction velocity $u^*$ deduced from PIV measurements in the liquid.
(a) in the absence of a grid; (b) with the G32 grid; (c) with the G64 grid.
Dashed lines correspond to fits based on the Schlichting law \eqref{eq:tauvsx}, with $C=0.029$ and and $x_0=-1530\,$mm (no-grid case), $C=0.032$ and $x_0=0\,$mm (G32 case) and $C=0.038$ and $x_0=150\,$mm (G64 case). The horizontal dotted line $u^* = 0.35~\mathrm{m~s^{-1}}$ corresponds to the critical friction velocity associated with the wrinkle-wave transition. The light grey area indicates the FS-SS measurement window.}
\label{fig:PIVutau1}
\end{figure} 

To determine the friction velocity $u^*$, we fit the velocity profile close to the surface with the parabolic law \eqref{eq:parab}. Then, differentiating with respect to $z$ and extrapolating to $z \rightarrow 0$ allows us to obtain $u^*$ using \eqref{eq:utau}. At large wind speed, light reflection on the wavy surface makes PIV measurements less reliable. This is why the fit is performed only in the range $|z|/h= 0.1-0.3$ in such cases. Figure \ref{fig:PIVutau1} shows the resulting streamwise variation of $u^*$ for various wind speeds $U_a$ and for the three different inlet conditions. A well-defined transition is observed for each inlet condition. Indeed, while $u^*$ slowly decreases with $x$ at low wind speeds (wrinkle regime), it exhibits a strongly non-monotonic behavior at larger wind speeds (wave regime).

\begin{figure}[tb]
\centering
\includegraphics[width=0.6\textwidth]{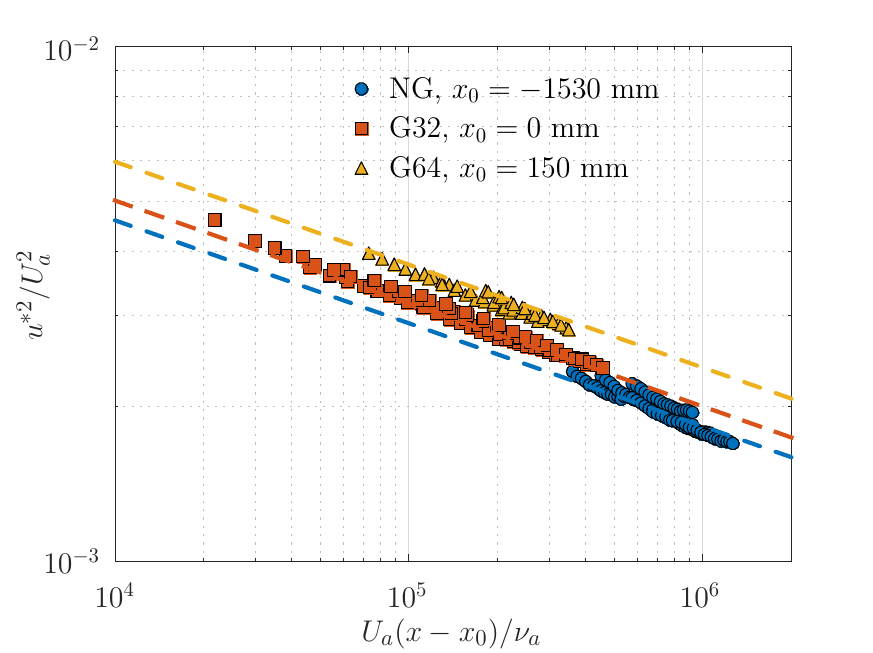}
\caption{Normalized friction velocity squared as a function of the Reynolds number $Re_x$ for wind speeds $U_a$ within the wrinkle regime. Dashed colored lines correspond to the Schlichting law \eqref{eq:tauvsx}, with a pre-factor $C =0.029$ (NG case), $C=0.032$ (G32) and $C=0.038$ (G64), respectively, and varying virtual origins $x_0$.}
\label{fig:Schlichting}
\end{figure}

Let us first discuss the low-speed regime. Here, the influence of free-stream turbulence on the friction velocity remains modest. Indeed, in the  window where the surface displacement measurements are performed, $u^*$ increases relative to the no-grid case by a factor of approximately $1.2$ with the grid G32 and $1.3$ with the grid G64  for a given wind velocity. This is to be compared to the huge increase of $\sqrt{\overline{u'^2}}$ which grows approximately by a factor of $7$ for case G32 and $14$ for case G64.  This weak increase of  $u^*$ indicates that the surface shear stress is primarily governed by the turbulence production in the boundary layer near the liquid surface rather than by the outer turbulence. In other words, the liquid surface is only weakly affected by the free-stream turbulence, which is efficiently screened by the boundary-layer turbulence. Nevertheless, despite its limited magnitude, we will see later that this increase in $u^*$ plays a crucial role in setting the wrinkle amplitude and controlling the transition from the wrinkle regime to the wave regime.

In this low-velocity regime, as Figs. \ref{fig:PIVutau1} and \ref{fig:Schlichting} make clear, the slow decay of $u^*$ with the streamwise position is well captured by the classical Schlichting law for a turbulent boundary layer developing over a rigid flat wall \citep{Schlichting}, namely
\begin{equation}
\frac{{u^*}^2}{U_a^2} = C Re_x^{-0.2} \,,
\label{eq:tauvsx}
\end{equation}
where $Re_x=U_a(x-x_0)/\nu_a$ is the Reynolds number based on the streamwise position, $x_0$ is a virtual origin, and $C$ is an empirical pre-factor. A best fit performed in the no-grid case  (Fig. \ref{fig:PIVutau1}a)  yields $C = 0.029$, in good agreement with the literature, with a large negative virtual origin $x_0=-1530$\,mm located far upstream from the liquid tank.  This agreement with the Schlichting law confirms that the wrinkle amplitude remains weak enough to leave the boundary layer in the air essentially unaltered. In the presence of grids, Figs. \ref{fig:PIVutau1}b and \ref{fig:PIVutau1}c reveal that not only do the values of $u^*$ exceed those found in the no-grid case, but they also exhibit a noticeably steeper decrease with $x$. The Schlichting law \eqref{eq:tauvsx} still captures this decrease, as shown in Fig.~\ref{fig:Schlichting} (red and yellow symbols), but with slightly larger pre-factors $C$ (approximately $ 0.032$ and $ 0.038$ for cases G32 and G64, respectively) and a shifted virtual origin ($x_0 = 0$~mm and $x_0=150$~mm, respectively for G32 and G64). The shift in $x_0$ suggests that the boundary layer is tripped well upstream of the liquid tank in the no-grid configuration, whereas in the presence of a grid, it develops at the beginning of the liquid tank in the G32 case, or slightly downstream in the G64 case.

Figure~\ref{fig:PIVutau1} shows that the monotonic decrease of $u^*$ with $x$ holds only up to a critical friction velocity $u_c^*$, which depends primarily on the wind speed $U_a$ but also on the fetch. At small fetch, this transition
occurs for $u^* \simeq 0.33-0.35$~\SI{}{\meter\per\second}, a value consistent with the threshold $u_c^* \simeq$ \SI{0.34}{\meter\per\second} reported by Aulnette {\it et al.}~\citep{aulnette2022kelvin} for the onset of waves in silicone oil with the same viscosity ($\nu_\ell =$ \SI{50e-6}{\meter\squared\per\second}), a slightly shallower depth ($h=35$~mm) and a weak free-stream turbulence intensity ($ T_u \simeq 2$\%). Beyond this threshold, $u^*$ exhibits a sharp increase at moderate fetches, due to the increasing effective surface roughness induced by the developing waves. A subsequent decrease of $u^*$ is also observed at larger fetch, with $u^*$ even dropping below $u_c^*$ in cases G32 and G64. As will be shown in the next section, this decay is associated with a decrease in the wave amplitude.

\begin{figure}[t]
\centering
\includegraphics[width=0.6\textwidth]{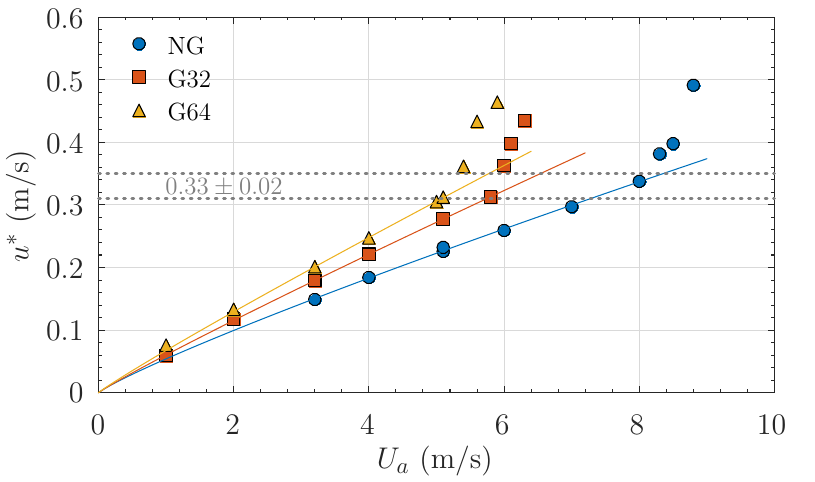}
\caption{Friction velocity $u^*$ measured at fetch $x=500$\,mm as a function of $U_a$ for the three inflow conditions. Solid lines are fits of the form $u^*=\alpha\,U_a^{p}$, with $p= 0.88$ for the NG case, and $p=0.94$ for the G32 and G64 cases.}  
\label{fig:UTevol2}
\end{figure}

Importantly, the critical friction velocity $u_c^*$ remains nearly constant across all inlet conditions, whereas the corresponding critical wind speed exhibits a strong dependence on these conditions. This effect is highlighted in Fig.~\ref{fig:UTevol2}, where the variation of $u^*$ with $U_a$ at a constant fetch $x=500\,$mm is shown. Below the transition, $u^*$ follows the Schlichting law \eqref{eq:tauvsx}, which may be recast in the simple form $u^* \propto U_a^{p}$, with $p \simeq 0.9 \pm 0.04$. In contrast, a strong increase of $u^*$ is observed when the friction velocity exceeds $u_c^*$. The critical wind speed at the onset of waves significantly decreases as the free-stream turbulence increases. Indeed, this transition occurs at $U_a \simeq$ \SI{8}{\meter\per\second} in the absence of a grid, while it occurs typically at \SI{6}{\meter\per\second} and \SI{5}{\meter\per\second} in cases G32 and G64, respectively. A more accurate estimate, based on the amplitude of the surface deformations, will be achieved in the next section.

\section{Surface deformations}
\label{sec:FSTeffect}

\subsection{Amplitude and characteristic lengths}

\begin{figure}[t]
\centering
\includegraphics[width=\textwidth]{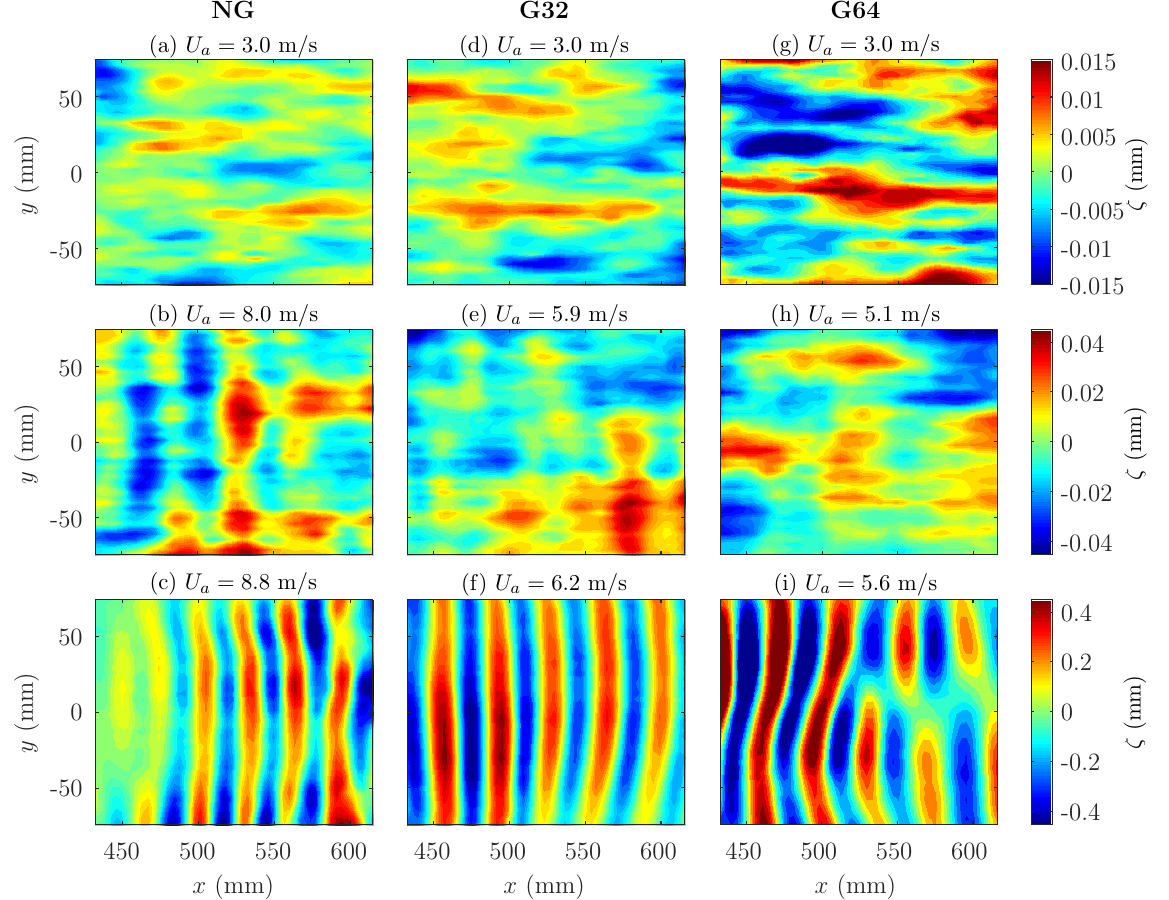}
\vspace{-2mm}
\caption{Instantaneous field of the surface height, $\zeta(x,y,t)$, in the absence of a grid (a,b,c), with grid G32 (d,e,f) and with grid G64 (g,h,i). Values of the wind speed are selected to illustrate the wrinkle regime (top), transitional regime (middle) and wave regime (bottom); note the different color scale used in each line.}
\label{fig:ISD}
\end{figure}

We now characterize the surface deformations induced by the wind for the different inlet conditions. Typical snapshots of the surface deformations measured by Free-Surface Synthetic Schlieren for each inflow configuration are presented in Fig. \ref{fig:ISD} for increasing wind speeds. All snapshots were recorded in the field of view $430\,$mm$\,\leq x\leq\,620\,$mm, including the position $x=500\,$mm at which friction velocities $u^*$ plotted in Fig.~\ref{fig:UTevol2} were measured.  For each inlet condition, a gradual variation of the surface pattern with increasing wind speed is observed:\vspace{1mm}\\
\indent 1. Below the critical velocity (Figs. \ref{fig:ISD}a, d, g), the surface exhibits disorganized, streamwise-elongated wrinkles with a typical amplitude of the order of \SI{10}{\micro\meter};\\
\indent 2. Near the onset of wave generation (Figs. \ref{fig:ISD}b, e, h), nearly two-dimensional waves with spanwise-oriented crests begin to overlap with the wrinkles, accompanied by a marked increase in the vertical displacements of the free surface which reach approximately \SI{30}{\micro\meter};\\
\indent 3. Above the critical velocity (Figs. \ref{fig:ISD}c, f, i), the surface deformation is dominated by regular, well-organized nearly two-dimensional waves with a characteristic wavelength of approximately $30-40\,$mm and amplitudes up to $0.5\,$mm.\vspace{1mm}\\
These properties are in qualitative agreement with the observations reported by Paquier {\it et al.} \citep{Paquier_2015} who used a wind tunnel without a grid, resulting in a free-stream turbulence level $ T_u \simeq 2\%$. Present experiments enlighten the two main effects of the free-stream turbulence intensity on the surface deformation: increasing $T_u$ slightly increases the amplitude of the wrinkles, and significantly decreases the critical wind speed for the onset of regular waves.\\
\begin{figure}[t]
\centering
\includegraphics[width=0.9\textwidth]{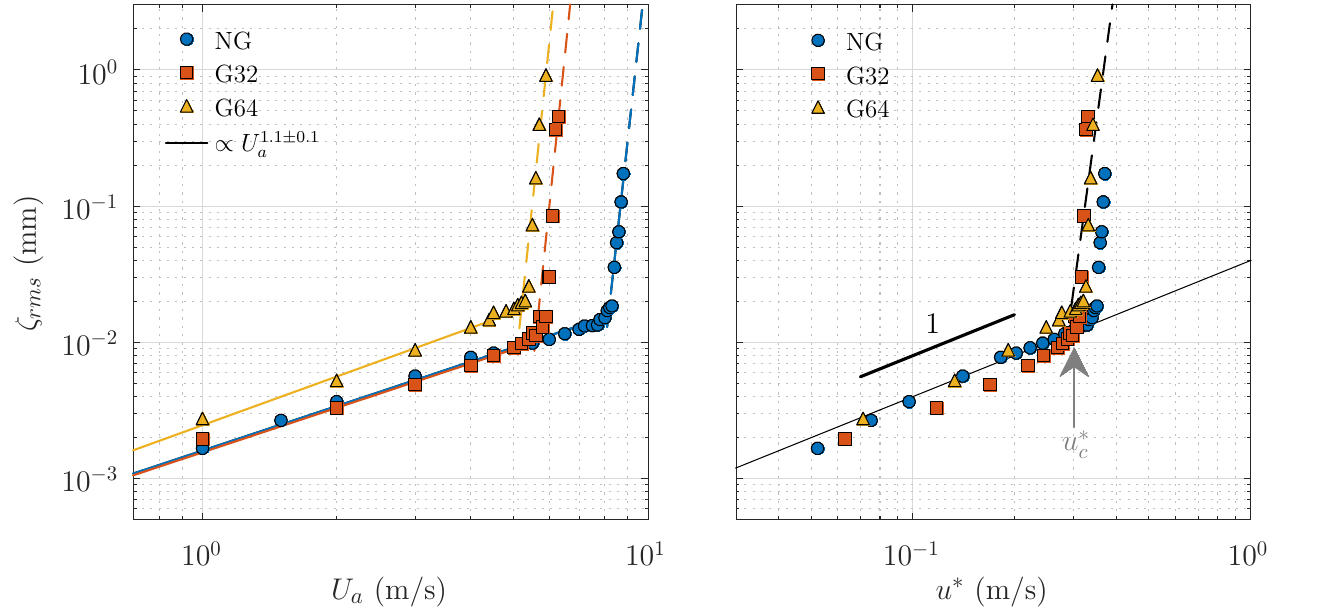}
\caption{Variation of the wave amplitude $\zeta_\text{rms}$ at $x=500$ mm as a function of: (a) the wind speed $U_a$; and (b) the local friction velocity $u^*$. The dashed lines in the wave regime correspond to power-law scalings shown as a guide to the eye.} 
\label{fig:Srms}
\end{figure}\indent
To quantify these effects, we determine the characteristic amplitude of the surface vertical displacements defined as
\begin{equation}
\zeta_\text{rms}(x) = {\Bigl\langle\overline{\zeta^2(x,y,t)}\Bigr\rangle}^{1/2}\,,
\end{equation}
where the averaging is performed first in time (overbar) and then along the spanwise direction (brackets). We first examine the dependence of $\zeta_\text{rms}$ on the wind speed and inlet configuration at a fixed fetch, $x=500\,$ mm; variations of $\zeta_\text{rms}$ with the fetch are analyzed in more detail in the next section.

Figure~\ref{fig:Srms} shows how $\zeta_{rms}$ varies as a function of the wind speed $U_a$ (Fig.~\ref{fig:Srms}a), and as a function of the local friction velocity $u^*$ (Fig.~\ref{fig:Srms}b). For the three inlet conditions, the curves reveal an almost linear increase in the wrinkle regime, followed by a much steeper increase in the wave regime. From the intersection of the extrapolated trends in each regime, we determine that the wrinkle-wave transition takes place at $U_{ac} = 8.0 \pm 0.1$ \:\mbox{m}\:\mbox{s}$^{-1}$ in the absence of a grid, and at $U_{ac}=5.9 \pm 0.1$ \:\mbox{m}\:\mbox{s}$^{-1}$, and $5.1 \pm 0.1$ \:\mbox{m}\:\mbox{s}$^{-1}$ in the G32 and G64 cases, respectively. 
These critical velocities closely match the change of slope of curves $u^*(U_a)$ in Fig.~\ref{fig:UTevol2}, confirming that the sudden increase in $u^*$ is directly associated with the wrinkle-wave transition. This sharp increase in $u^*$ beyond the transition highlights the fact that, by increasing the surface roughness, waves strongly increase the surface stress at the interface.

As expected from the above discussion, when $\zeta_{rms}$ is plotted as a function of $u^*$ (Fig. \ref{fig:Srms}b), the curves corresponding to the three inlet conditions approximately converge. This confirms that the transition takes place at $u^*_{c} \simeq 0.33 \pm 0.02$ \SI{}{\meter\per\second}, corresponding to a characteristic surface deformation of $\zeta_{rms} \simeq 16\, \pm\,$\SI{4}{\micro\meter}. Below the transition, in the wrinkle regime, a power-law fit of the form $\zeta_{rms} \propto {u^*}^n$ yields an exponent $n \simeq 1.1 \pm 0.1$. Although a quantitative comparison with theory is difficult owing to the limited range of $u^*$ accessible in the wrinkle regime, this exponent is significantly smaller than the theoretical prediction $n=1.5$ of \cite{Perrard2019} summarized in Eq.~\eqref{eq.perrard}. However, the measured exponent is consistent with the experimental results of Paquier \textit{et al.} \citep{Paquier_2016} for liquid viscosities close to that used here, where power-law fits for $\nu_\ell=30\times10^{-6}$~m$^2$s$^{-1}$ and $85\times10^{-6}$~m$^2$s$^{-1}$ yield exponents in the range $n\simeq1.1$–$1.2$. This quasi-linear variation of $\zeta$ therefore appears to be a robust feature of the wrinkle regime.

The decrease of the critical wind speed $U_{ac}$ in the presence of a grid may partly result from the overshoot observed in the $\overline{u}(z)$ profiles in the boundary-layer region (see Fig. \ref{fig:HWprofiles}). If the wrinkle–wave transition were controlled by the wind velocity at the outer edge of the boundary layer, this overshoot would indeed trigger the transition at a lower free-stream velocity. However, the overspeed observed in Fig.~\ref{fig:HWprofiles} reaches only 15–20\%, whereas the reduction in $U_{ac}$ is more pronounced, typically 25–35\% compared to the no-grid configuration. This difference indicates that the influence of the grid on the transition does not reduce to a mere shift in the effective wind speed, but also involves modifications of the turbulent activity within the boundary layer. The marked reduction in $U_{ac}$ when a grid is present suggests that the extra turbulent intensity induced by the grid results in an increase of the wrinkle amplitude, which in turn allows the wrinkle-wave transition to take place at a lower wind speed.

The fact that the wrinkle amplitude is approximately constant at the transition ($\zeta_{rms}\simeq 16 \pm 4~\mu$m) is in qualitative agreement with the criterion proposed by Perrard {\it et al.}~\citep{Perrard2019}, which states that the wrinkle-wave transition occurs when $\zeta_{rms}$ reaches a certain fraction of the viscous scale $\delta_\nu = \nu_a / u^*$. Here, one has $\delta_\nu \simeq 45 \pm 5$ \SI{}{\micro\meter} at the transition, yielding a ratio $\zeta_{rms} / \delta_\nu \approx 0.36\pm0.1$. This value is significantly larger than that found in the experiments of Paquier~{\it et al.} \citep{Paquier_2016} ($\zeta_{rms} / \delta_\nu \simeq 0.11$) or in the simulations of Li and Shen \cite{LiShen2022}. The difference may result from several factors. First of all, the experiments of \citep{Paquier_2016} were performed in a water–glycerol mixture having a surface tension $\gamma \simeq 0.06$ N~m$^{-1}$, significantly larger than that of the silicone oil used here ($\gamma = 0.021$~N~m$^{-1}$). This lower surface tension, together with the slightly thicker liquid layer used here, may contribute to increase significantly the wrinkle amplitude for a given level of pressure fluctuations. Indeed, a lower $\gamma$ makes surface deformation easier, contributing to increase the interface roughness at a given wind speed, as shown in the wave regime by Matsuda~{\it et al.} \citep{Matsuda_2023}. In addition, pressure fluctuations in the turbulent boundary layer may themselves be highly sensitive to details of the wind-tunnel geometry, in particular the channel height (500~mm here, compared to 105~mm in \citep{Paquier_2016}). Although these differences make it difficult to quantitatively compare the amplitude of the wrinkles in the two experiments, the close agreement in the critical friction velocity $u^*_c$ strongly suggests that this parameter provides a robust criterion for describing the transition between wrinkles and waves.

The influence of the inlet conditions is not limited to the wrinkle amplitude, but may also affect their geometry. Wrinkles can be seen as transient wakes produced by turbulent pressure fluctuations travelling in the boundary layer.
Their characteristic width is essentially governed by the width of the pressure patches, which scales like the boundary layer thickness, while their length is governed by the wake dynamics of these moving pressure sources~\citep{Perrard2019,Nove_2020}. To quantify the characteristic shape of surface deformations, we compute the  two-point correlation function
\begin{equation}
    C(\mathbf{r}) = 
    \frac{\left< \zeta(\mathbf{x},t) \zeta(\mathbf{x+r},t) \right>}
    {\left< \zeta(\mathbf{x},t)^2 \right> }\,,
\end{equation}
where $\mathbf{r}=r_x\mathbf{e_x}+r_y\mathbf{e_y}$, and $\left<.\right>$ denotes the average over the entire two-dimensional field of view $(x,y)$ and over time. From this, we define the longitudinal and transverse correlation lengths $\Lambda_i$ (with $i = x,y$) as $\Lambda_i=6r_{i0}$, with $r_{i0}$ being the location of the first zero-crossing of $C(r_i)$. With this definition, one gets $\Lambda_i = \lambda$ in the case of a monochromatic wave field of wavelength $\lambda$ propagating in the $\mathbf{e}_i$-direction. In the case of disorganized wrinkles, $\Lambda_x$ and $\Lambda_y$ still provide characteristic wavelengths of the deformation pattern in the longitudinal and transverse directions.

\begin{figure}[t]
\centering
\includegraphics[trim=0 0 0 0, clip=, height=0.35\textwidth]{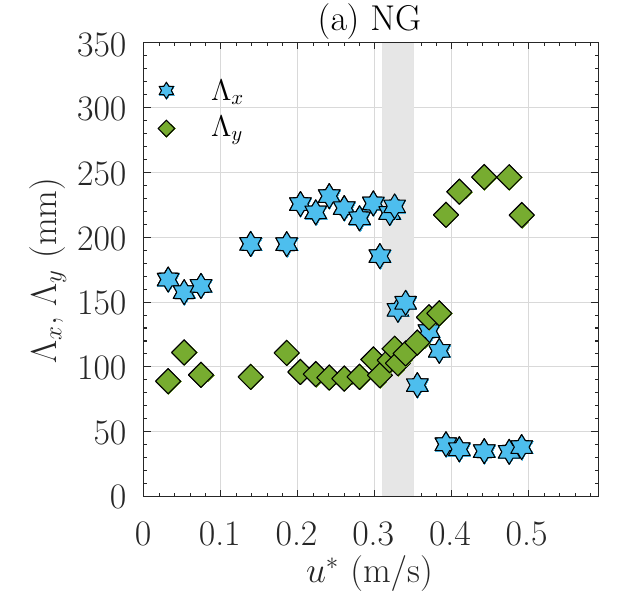}
\includegraphics[trim=61 0 0 0, clip=, height=0.35\textwidth]{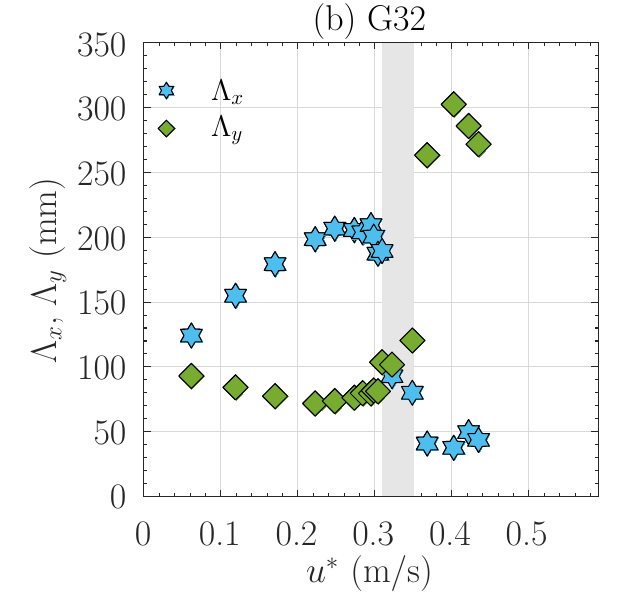}
\includegraphics[trim=61 0 0 0, clip=, height=0.35\textwidth]{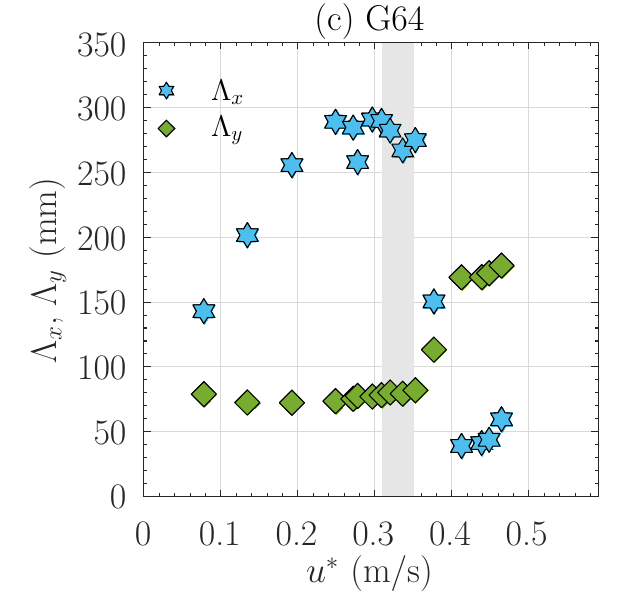}
\caption{Longitudinal (blue stars) and transverse (green diamonds) correlation lengths $\Lambda_x$ and $\Lambda_y$, averaged over the window $x \in [430, 620]\,$mm, as a function of friction velocity $u^*$ for (a) the NG case, (b) the G32 case, and (c) the G64 case. The vertical grey strip corresponds to the critical friction velocity $u^*_c=0.33\pm0.02$~\SI{}{\meter\per\second}.}
\label{fig:Corr2D} 
\end{figure}

Figure \ref{fig:Corr2D} shows the variation of the two correlation lengths as a function of the friction velocity for the three inflow conditions. All curves exhibit a marked crossover of  $\Lambda_x$ and $\Lambda_y$ at $u^* \simeq u^*_{c}$,  a hallmark  of the geometrical transition from disorganized wrinkles to nearly two-dimensional waves. Below the transition, wrinkles are characterized by a nearly constant characteristic width $\Lambda_y \simeq 70-100\,$mm, and a length $\Lambda_x$ increasing with $u^*$ from approximately 150 to 300\,mm. This increase in $\Lambda_x$ is consistent with the narrower wake generated by a pressure perturbation as its convection velocity increases~\citep{rabaud2013ship,Nove_2020}. Beyond the transition, the longitudinal correlation length decreases significantly to $\Lambda_x  \simeq 40\,$mm, the typical wavelength of the regular waves, while the transverse length $\Lambda_y$ strongly increases  ($\Lambda_y$ would be infinite for purely two-dimensional waves). The influence of the inlet conditions on $\Lambda_x$ and $\Lambda_y$ is complex but not major. We notice that, in the wrinkle stage, $\Lambda_x$ increases more markedly with $u^*$ in the presence of grids, reaching its largest value in the case of the G64 grid. This may be the consequence of a larger spatial extension of pressure fluctuations resulting from the larger mesh size. Also, beyond the transition, $\Lambda_y$ is seen to be significantly smaller in the G64 case compared to the G32 case. This may result from a stronger spanwise modulation of the velocity, affecting the  two-dimensionality of the wave field.

\subsection{Variation with fetch}

The amplitude of surface deformations depends on the wind forcing, but also on the fetch at which measurements are performed. This dependence with respect to $x$ is already visible in the bottom row of Fig. \ref{fig:ISD}, with a wave amplitude increasing with $x$ in the no-grid case, while this amplitude stays nearly constant or even slightly decreases in the G32 and G64 cases. This decrease of $\zeta_{rms}$ with $x$ is further confirmed by visual inspection. Indeed, surface waves are seen to be strongly attenuated, or even to vanish, near the downstream end of the liquid tank for wind velocities slightly beyond the transition.

\begin{figure}[t]
\centering
\includegraphics[trim={0 0 0 0}, clip, height=0.33\textwidth]{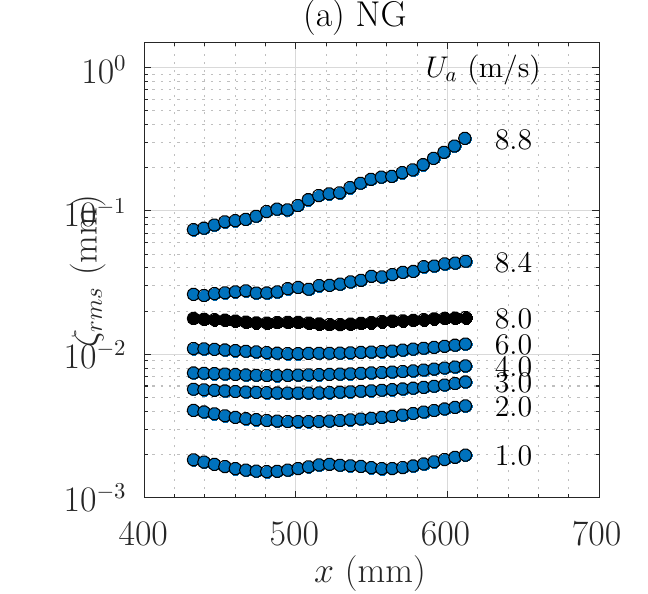}
\includegraphics[trim={57 0 0 0}, clip, height=0.33\textwidth]{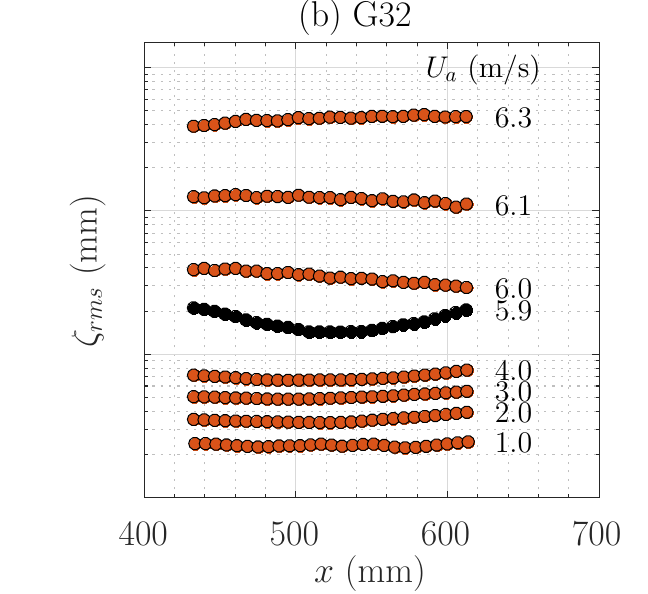}
\includegraphics[trim={57 0 0 0}, clip, height=0.33\textwidth]{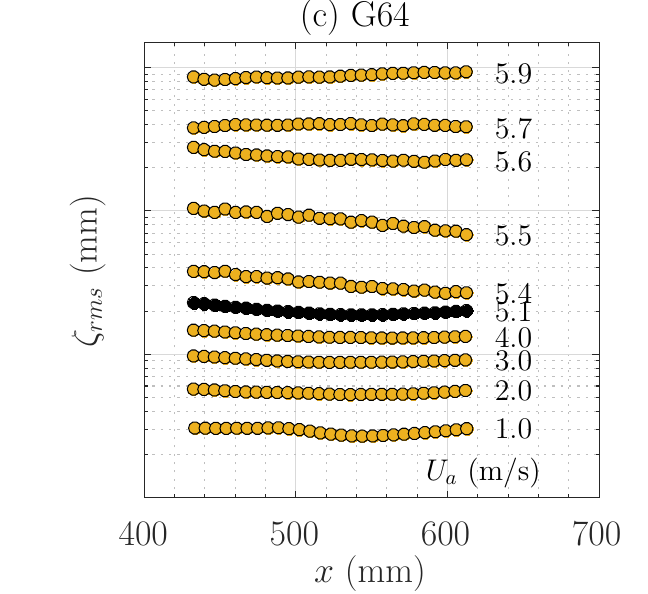}
\caption{Linear-log representation of the longitudinal variation of the amplitude of surface deformations for various wind speeds. (a) No-grid case, (b) with grid G32, and (c) with grid G64. Data associated with black bullets are obtained right at the wrinkle-wave transition, \textit{i.e.}, at $U_a = U_{ac}$.
}
\label{fig:SrmsX}
\end{figure}

Influence of the fetch on surface deformations is enlightened in Fig. \ref{fig:SrmsX} which displays the longitudinal variation of $\zeta_\text{rms}(x)$ for the three inlet conditions at various wind speeds, including right at the transition (black bullets). 
In the wrinkle regime, $\zeta_\text{rms}$ is seen to be almost independent of the fetch for all inlet conditions, at least over the narrow range of $x$ where FS-SS measurements are performed. In contrast, in the wave regime, $\zeta_\text{rms}$ exhibits a more complex behavior, with either a growth or a decay for increasing fetches. It is reasonable to hypothesize that this non-monotonic behavior is related to the non-monotonic variation of $u^*$ with $x$ evidenced in Fig. \ref{fig:PIVutau1}. To elaborate more on this connection, it is necessary to distinguish between wrinkles and waves.

In the wrinkle regime, $\zeta_{rms}(x)$ correspond to the {\it passive} response of the liquid surface to the turbulent forcing exerted by the wind~\citep{Perrard2019}. This is because $\zeta_{\rm rms}$ remains small ($\simeq 1-10$~\SI{}{\micro\meter}) compared to the viscous length scale $\delta_\nu = \nu / u^*$ ($ 45-300$~\SI{}{\micro\meter}), so that surface deformations do not modify significantly the air boundary layer. In this regime, one therefore expects an essentially {\it local} relationship between $\zeta_{rms}(x)$ and pressure fluctuations $p_{rms}(x)$ within the boundary layer. However, the link between $p_{rms}$ and $u^*(x)$ is not as straightforward, as it depends on the detailed structure of the boundary layer, which itself is influenced by the free-stream turbulence. Seeking only an order-of-magnitude estimate, one may write $p_{rms}(x) \simeq p^+ \rho_a {u^*}^2(x)$, with $p^+$ a dimensionless pre-factor depending on the Reynolds number and inlet conditions. For a standard turbulent boundary layer with no free-stream turbulence, one has $p^+ \simeq 2 \pm 0.5$ in the range of Reynolds numbers considered here \citep{hu2006wall, jimenez2008turbulent}, but larger values are expected in the presence of grids. Owing to the weak decay of $u^*$  with $x$ over the limited range where surface measurements are performed (less than 2\% in the no-grid case and 5\% with grids according to Fig. \ref{fig:PIVutau1}), $p_{rms}$ is expected to be essentially independent of $x$, which directly translates into nearly-uniform longitudinal profiles of $\zeta_{\rm rms}$, in agreement with the almost flat profiles observed in Fig. \ref{fig:SrmsX}. The remaining scatter in the surface deformation observed in Fig.~\ref{fig:Srms}b when $\zeta_{\rm rms}$ is plotted as a function of $u^*$ at constant $x$ likely reflects the fact that $p_{\rm rms}$ depends not only on $u^*$, but also on the detailed structure of the boundary layer, hence on the inlet conditions.

In the wave regime, the situation is generally more intricate because $\zeta_{\mathrm{rms}}(x)$ and $u^*(x)$ become interdependent quantities coupled through non-local feedback mechanisms. Indeed, the amplitude $\zeta_{\mathrm{rms}}$ of surface deformations at a given fetch results from the cumulative action of the wind forcing over the entire region extending from $x=0$ to the current longitudinal position. This process is dominated by the amplification of waves through the Miles mechanism, partly counterbalanced by viscous dissipation in the bulk of the liquid \citep{Miles_1957,Miles1962generation,Gastel1985phase,Zhang_2023}. Conversely, by modifying the surface roughness, waves enhance the turbulence in the air boundary layer and thus increase $u^*$ \citep{Miles1967generation,Miles1993surface}.
The high viscosity of the liquid may also partly simplify the general picture because the large dissipation in the liquid reduces the cumulative effect of the wind forcing, leading to a more local relationship between $u^*$ and $\zeta_{\rm rms}$. This enhanced locality may help connect the non-monotonic longitudinal variation of $\zeta_{\rm rms}$ revealed by Fig. \ref{fig:SrmsX} with the non-monotonic variation of $u^*$ evidenced by Fig.~\ref{fig:PIVutau1}. This is the purpose of the simple model developed in the next section.

\begin{figure}[t]
\centering
\includegraphics[trim=0 0 0 0, clip=, height=0.33\textwidth]{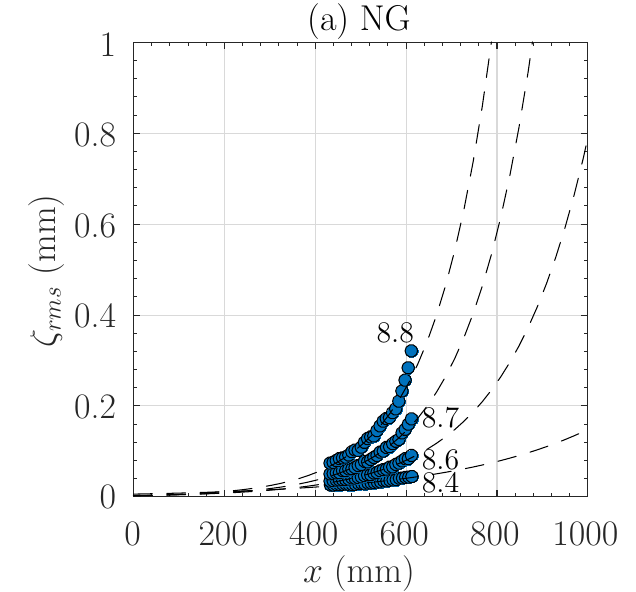}
\includegraphics[trim=57 0 0 0, clip=,  height=0.33\textwidth]{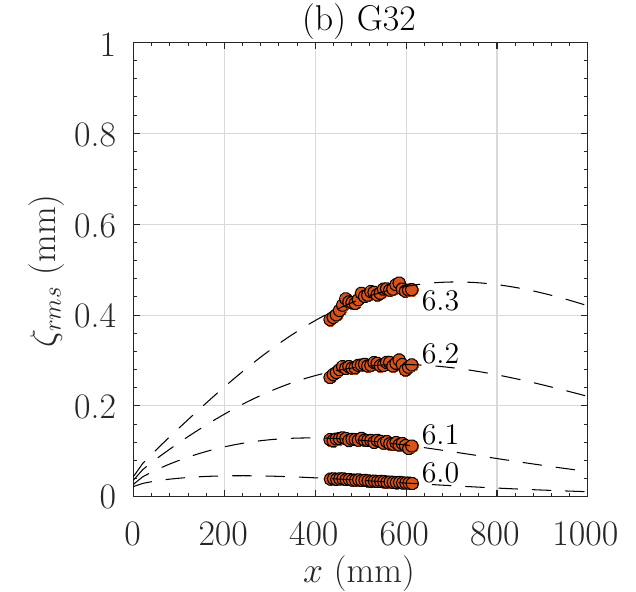}
\includegraphics[trim=57 0 0 0, clip=, height=0.33\textwidth]{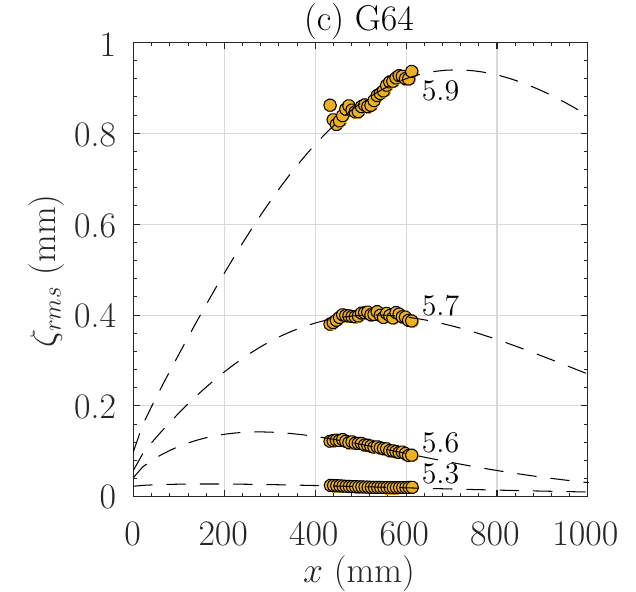}
\caption{Streamwise evolution of the wave amplitude $\zeta_{rms}$ for wind speeds beyond the critical velocity $U_{ac}$.  Dashed lines are fits based on the model \eqref{eq:fitzrms}.
}
\label{fig:SrmsX3}
\end{figure}

\subsection{A simple model for the non-monotonic variation of the wave amplitude with fetch}

To rationalize the non-monotonic variation of $\zeta_{rms}(x)$ with the fetch   in the wave regime, we introduce the following simplified model. We consider that the wave field is dominated by a single Fourier component with a wavelength $\lambda \simeq 40\,$mm propagating in the $x$-direction. Since this wavelength is significantly larger than the capillary-gravity wavelength $\lambda_c \simeq 9.3\,$mm, it belongs to the gravity regime, so that the wave energy density per unit surface is $E \simeq \rho_\ell g \zeta_{rms}^2$. Since the characteristic wave slope $k\zeta=2\pi\zeta/\lambda$ remains moderate close to the onset of waves ($k\zeta < 0.08$ for $\zeta$ up to  0.5\,mm), we assume a purely linear dynamics and disregard the retroaction of the wave field on $u^*$. In the statistically stationary state, the evolution of the wave energy density is governed by~\citep{Kahma_1988,grare2013growth,PEIRSON:2008}
\begin{equation}
(c_g+U_s) \frac{\partial}{\partial x} E = P - D,
\label{eq:EPD}
\end{equation}
with $c_g(k)$ the group velocity of the dominant wave component, $U_s$ the surface drift current, $P(k)$ the rate at which wind supplies energy to the wave field and $D(k)$ the rate at which the wave energy is dissipated in the liquid by viscosity. Here, the wave transport by the group velocity is the dominant contribution: one has $c_g \simeq 0.14$~\SI{}{\meter\per\second} for $\lambda\simeq 40$~mm, while $U_s < 0.05$~\SI{}{\meter\per\second} (see Fig.~\ref{fig:PIVumean}).
Assuming that, close to the onset, waves grow exponentially due to the wind-induced stress (pressure and viscous shear stress) applied on the liquid surface \citep{Miles_1957,Miles1962generation}, the energy injection rate may be written in the form $P = \gamma E$, with the temporal growth rate $\gamma$ given by~\citep{Miles_1957,Miles1959_part2,plant1982relationship}
\begin{equation}
   \gamma = \beta \omega \frac{\rho_a}{\rho_\ell} \left( \frac{u^*}{c} \right)^2,
\label{eq:Miles} 
\end{equation}
where $\omega(k)$ denotes the characteristic radian frequency, $c=(g/k)^{1/2}$ is the corresponding phase velocity, and $\beta$ is a non-dimensional parameter. Compilation of field and laboratory data obtained for wind-generated waves over water suggests an average value $\beta \simeq 32\pm 16$~\citep{plant1982relationship,Janssen_2004}, while recent experiments using silicone oil with a kinematic viscosity $\nu_l=$ \SI{50e-6}{\meter\squared\per\second} provide comparable values, $\beta \simeq 45 \pm 20$~\citep{Zhang_2023}.

The dissipation term $D$ includes bulk dissipation, $D(k) = - 4 \nu_\ell k^2 E$ \citep{Lamb}, and dissipation in the surface and bottom boundary layers. Given the values of $\nu_l$, $k$, and the liquid depth considered here, the bulk contribution is dominant~\citep{Zhang_2023}. We can therefore rewrite \eqref{eq:EPD} in the approximate form
\begin{equation}
(c_g+U_s) \frac{\partial}{\partial x} E \approx \left( \beta\frac{ \rho_a}{\rho_\ell } u^{*2}(x) - 4 \nu_\ell\omega \right) \frac{ \omega E}{c^2}\,,
\label{eq:EPDc}
\end{equation}
which defines the critical friction velocity $u_c^*$ at which the net growth rate vanishes.
If $u^*$ were $x$-independent, Eq.~\eqref{eq:EPDc} would predict an exponential decrease (for $u^* < u^*_c$) or increase (for $u^* > u^*_c$) of the wave amplitude with $x$. However, due to the spatial decay of $u^*$ in the streamwise direction, the net growth rate changes sign at some critical fetch, $x_m$, resulting in a non-monotonic wave amplitude. This is consistent with Figs.~\ref{fig:PIVutau1} and \ref{fig:SrmsX}. Indeed, beyond the wave onset, $u^*$ remains consistently larger than $u^*_c$ over the available range of fetches in the absence of grids, leading to a positive growth rate and, consequently, to a monotonic increase of $\zeta_{rms}$ with $x$. In contrast, in the presence of grids, the longitudinal $u^*$-profiles drop below the threshold value $u^*=u^*_c$ beyond a certain fetch $x_m$ that decreases from $1000\,$mm to $600\,$mm as the wind speed increases, consistent with the observed longitudinal variation of $\zeta_{rms}$ in the measurement window.

Further prediction of the evolution of the wave energy with the fetch is difficult, as the feedback of the waves on the friction velocity is not known. An approximate model can, however, be obtained by neglecting this feedback and extrapolating the Schlichting relation \eqref{eq:tauvsx} beyond the critical value $u^*_c$. With these crude assumptions, Eq.~\eqref{eq:EPDc} predicts that the wave amplitude reaches its maximum at a fetch $x_m$ such that 
\begin{equation}
x_m \approx x_0 + \frac{\nu_a}{U_a} \left( \frac{C \beta}{4}\frac{ \rho_a}{ \rho_\ell}\frac{ U_a^2}{\nu_\ell \omega} \right)^5\,,
\label{eq:xm}
\end{equation}
with $C$ the constant involved in Eq.~\eqref{eq:tauvsx}. It is noteworthy that the dependence of $x_m$ on the wind speed is extremely steep ($\propto U_a^9$), in qualitative agreement with observations. Indeed, in the absence of a grid (large $U_a$), $x_m$ likely exceeds the tank length, whereas in the presence of a grid (smaller $U_a$), a shorter $x_m$ is obtained, which shifts significantly downstream as $U_a$ increases. Integrating Eq.~\eqref{eq:EPDc} with $u^*(x)$ obeying the Schlichting relation \eqref{eq:tauvsx} yields a non-monotonic variation of $\zeta_{rms}(x)$ in the form
\begin{equation}
\zeta_{rms}(x) \propto \exp \left\{ A (x-x_0)^{4/5} \left[ \frac{5}{4} - \left(\frac{x-x_0}{x_m-x_0} \right)^{1/5} \right]\right\},
\label{eq:fitzrms}
\end{equation}
where the dimensional coefficient 
\begin{equation}
A=C\beta\frac{\rho_a}{\rho_\ell}\frac{\omega U_a^2}{(c_g+U_s)c^2}\left(\frac{\nu_a}{U_a}\right)^{1/5}
\end{equation}
 depends on the wind speed and wave properties. Data in Fig.~\ref{fig:SrmsX3} are consistent with this behavior, both with and without grids: whereas $\zeta_{rms}$ does not exhibit a maximum in the no-grid configuration, it displays a clear maximum in the G32 and G64 cases at a fetch $x_m$ that shifts downstream as the wind velocity increases. 

Such a non-monotonic variation of the wave amplitude is specific to the high liquid viscosity used in the present experiments, and is unlikely to extend to smaller viscosities. In experiments performed in water, the viscous attenuation is much smaller, so the fetch at which the wave amplitude reaches its maximum ($x_m\propto \nu_\ell^{-5}$ according to \eqref{eq:xm}) would be shifted to considerably larger streamwise positions. In this case, other effects such as wave saturation by nonlinear interactions are expected to take place much earlier, making the above predictions for $\zeta_{rms}(x)$ irrelevant. Conversely, with a much more viscous liquid, $x_m$ would be considerably smaller, so that waves would be visible only near the entrance of the tank. This is indeed consistent with experiments performed with highly viscous silicone oils having kinematic viscosities up to \SI{5000e-6}{\meter\squared\per\second}~\citep{aulnette2022kelvin}, in which periodic waves are only observed at small fetch.


\section{Summary and final comments}

We carried out experiments aimed at characterizing how free-stream turbulence in the air affects the generation of surface waves by wind over a viscous liquid. Using two different turbulence-generating grids inserted downstream of the converging section of a low-turbulence wind tunnel, we could increase the turbulence intensity from 0.6\% in the absence of a grid to roughly 4 and 8\% at the location of the free-surface measurements. Using a variety of complementary measurement techniques, we could then quantify the influence of the inlet conditions and of the resulting non-standard turbulent boundary layer on the characteristics of the wrinkles (weak-amplitude turbulence-driven surface deformations) at low wind speed and the onset of regular nearly two-dimensional waves at larger wind speed. Our main findings are as follows:\vspace{1mm}\\
\indent 1. For all inlet conditions, surface deformations exhibit a clear transition from the wrinkle regime to the regular wave regime as the wind speed $U_a$ is increased, consistent with earlier observations carried out with low free-stream turbulence intensities \citep{Paquier_2015,Paquier_2016};\\
\indent 2. For a given wind speed, the grids enhance the friction velocity $u^*$  by about 20–30\% in the region of interest. In the wrinkle regime, the longitudinal decay of $u^*$ with the fetch is similar to that observed in a standard boundary layer over a rigid flat wall with negligible free-stream turbulence (Schlichting law). Nevertheless, the pre-factor and virtual origin involved in this empirical law depend on the inlet condition;\\
\indent 3. The grids significantly reduce the critical wind speed corresponding to the onset of regular waves, from $U_{ac} \simeq 8$~\SI{}{\meter\per\second} to 6 and 5~\SI{}{\meter\per\second}. However, the critical friction velocity at the transition remains almost constant, $u_c^* \simeq 0.33 \pm 0.02$~\SI{}{\meter\per\second}, in fairly good agreement with previous studies employing liquids with comparable viscosities;\\
\indent 4. Beyond the critical wind speed, both the local friction velocity $u^*(x)$ and the local wave amplitude $\zeta_{rms}(x)$ strongly increase with $U_a$ (or $u^*$). Yet, the functional relationship $\zeta_{rms} = f(u^*)$ at a given fetch $x$ remains unchanged for all inlet conditions;\\
\indent 5. In the wave regime, a non-monotonic variation of $u^*(x)$ and $\zeta_{rms}(x)$ with fetch is observed. A simple model based on the governing equation for the wave energy density helps rationalize this behavior by considering the joint effect of the wind-driven input governing the wave growth (which itself is modulated by the longitudinal variations of $u^*(x)$) and of the viscous dissipation of wave energy in the bulk.\vspace{1mm}\\ 
\indent To date, most models, experiments and numerical simulations of wind-generated waves consider a canonical turbulent boundary layer in the air flow above the free surface. 
Changing the properties of this boundary layer provides a way to quantify the relationship between the turbulent forcing and the response of the surface, and to explore the transition between the Phillips mechanism~\citep{Phillips_1957}, assumed to drive the generation of wrinkles, and the Miles mechanism~\citep{Miles_1957}, considered to govern the wave growth. At low wind speed, surface deformations are driven by the forcing of the surface by turbulent pressure fluctuations in the boundary layer. At larger wind speed, the wave growth predicted by the Miles model is primarily governed by the mean velocity profile in the air, which itself depends on the inner structure of the boundary layer. Both mechanisms are therefore expected to depend on the details of the boundary layer and the turbulent activity therein.

A natural extension of the present study, currently underway, involves replacing the passive grids used here with an active grid that allows much higher levels of free-stream turbulence (up to $\simeq20\%$). This will enable to determine whether or not the trends revealed by the present study for $T_u\leq8\%$, especially the fact that the wrinkle-wave transition takes place at a constant $u^*$ irrespective of $T_u$, still hold in the presence of much more intense free-stream turbulence.

\begin{acknowledgments}

We thank Jean-Dominique Barron and Christophe Korbuly for designing and implementing the experiment. 
We are grateful to Yara Abidine and Moïse Marchal for their assistance with the measurements. 
This work was supported by the project ``ViscousWindWaves'' of the French National Research Agency under grant ANR-18-CE30-0003.

\end{acknowledgments}

\bibliography{Biblio_Viscous_Waves}

@article{LiShen2022,
	author = {Tianyi Li and Lian Shen},
	journal = {J. Fluid Mech.},
	pages = {A41},
	title = {The principal stage in wind-wave generation},
	volume = {934},
	doi={https://doi.org/10.1017/jfm.2021.1153},
	year = {2022}}

@article{LiShen2025,
	author = {Tianyi Li and Lian Shen},
	journal = {J. Fluid Mech.},
	pages = {A8},
	title = {A theoretical study of the upper bound of surface elevation variance in the {P}hillips initial stage during wind-wave generation},
	volume = {1006},
	doi={https://doi.org/10.1017/jfm.2025.13},
	year = {2025}}

@article{Geva2022,
	author = {Geva, M. and Shemer, L.},
	journal = {Phys. Rev. Lett.},
	pages = {124501},
	title = {Excitation of initial waves by wind: A theoretical model and its experimental verification},
	volume = {128},
	year = {2022},
	doi={https://doi.org/10.1103/PhysRevLett.128.124501}}

@article{Kumar2024,
	author = {Kumar, K. and Shemer, L.},
	journal = {J. Fluid Mech.},
	pages = {A22},
	title = {Spatial growth rates of young wind waves under steady wind forcing},
	volume = {984},
	year = {2024},
	doi={https://doi.org/10.1017/jfm.2024.228}}

@article{Ishimura2025,
	author = {Ishimura, M. and Mergui, S. and Ruyer-Quil, C. and Dietze, G.F.},
	journal = {J.  Fluid Mech.},
	pages = {A30},
	title = {A new upward-convective short-wave instability mode in gas-sheared falling liquid films},
	volume = {1024},
	doi={https://doi.org/10.1017/jfm.2025.10889},
	year = {2025}}

@article{ayet2022,
	author = {Ayet, Alex and Chapron, Bertrand},
	journal = {Bound.-Layer Meteor.},
	pages = {1–33},
	title = {The Dynamical Coupling of Wind-Waves and Atmospheric
Turbulence: A Review of Theoretical and Phenomenological
Models},
	volume = {183},
	year = {2022},
	doi={https://doi.org/10.1007/s10546-021-00666-6}}

@article{chaubet2024effect,
	author = {Chaubet, C and Kern, N and Manna, MA},
	journal = {Phys. Fluids},
	title = {Effect of viscosity on wind-driven gravitation waves},
	volume = {36},
	pages={092109},
	year = {2024},
	url={https://doi.org/10.1063/5.0221941}}

@article{Matsuda_2023,
	author = {Matsuda, K and Komori, S and Takagaki, N and Onishi, R},
	journal = {J. Fluid Mech.},
	title = {Effects of surface tension reduction on wind-wave growth and air-water scalar transfer},
	volume = {960},
	pages={A22},
	year = {2023},
	doi={https://doi.org/10.1017/jfm.2023.144}}

@article{Zhang_2023,
  title = {Wind-wave growth over a viscous liquid},
  author = {Zhang, J. and Hector, A. and Rabaud, M. and Moisy, F.},
  journal = {Phys. Rev. Fluids},
  volume = {8},
  issue = {10},
  pages = {104801},
  numpages = {22},
  year = {2023},
  publisher = {American Physical Society},
  doi = {10.1103/PhysRevFluids.8.104801},
  url = {https://link.aps.org/doi/10.1103/PhysRevFluids.8.104801}
}

@article{bourguet23_JFS,
title = {A wind tunnel investigation of the effects of end and laminar/turbulent inflow conditions on cylinder vortex-induced vibrations},
journal = {J. Fluids Struct.},
volume = {123},
pages = {104015},
year = {2023},
issn = {0889-9746},
doi = {https://doi.org/10.1016/j.jfluidstructs.2023.104015},
author = {Bourguet, R. and Mathis, R.}
}

@article{cardesa12_EiF,
	author = {J.~I. Cardesa and T.~B. Nickels and J.R. Dawson},
	journal = {Exp. Fluids},
	pages = {1611--1627},
	doi = {10.1007/s00348-012-1278-4},
	title = {2{D PIV} measurements in the near field of grid turbulence using stitched fields from multiple cameras},
	volume = {52},
	year = {2012}
}

@book{Schlichting,
	author = {H. Schlichting},
	edition = {8th},
	publisher = {Springer},
	title = {Boundary Layer Theory},
	year = {2000}}

@book{spurk08_BOOK,
	author = {Spurk, J. H. and Aksel, N.},
	edition = {2nd},
	publisher = {Springer},
	title = {Fluid Mechanics},
	year = {2008}}

@article{owen57_JFM,
	author = {Owen, P. R. and Zienkiewicz, H. K.},
	doi = {10.1017/S0022112057000336},
	journal = {J. Fluid Mech.},
	pages = {521--531},
	title = {The production of uniform shear flow in a wind tunnel},
	volume = {2},
	year = {1957}}

@article{lau68_JFM,
	author = {Lau, Y. L. and Baines, W. D.},
	doi = {10.1017/S0022112068001643},
	journal = {J. Fluid Mech.},
	pages = {721--738},
	title = {Flow of stratified fluid through curved screens},
	volume = {33},
	year = {1968}}

@article{mehta85_AIAAj,
	author = {Mehta, R. D.},
	doi = {10.2514/3.9089},
	journal = {AIAA J.},
	pages = {1335--1342},
	title = {Turbulent boundary layer perturbed by a screen},
	volume = {23},
	year = {1985}}

@article{mohamed90_JFM,
	author = {Mohamed, M. S. and LaRue, J. C.},
	journal = {J. Fluid Mech.},
	pages = {195--214},
	title = {The decay power law in grid-generated turbulence},
	volume = {219},
	doi={https://doi.org/10.1017/S0022112090002919},
	year = {1990}}

@article{lavoie07_JFM,
	author = {Lavoie, P. and Djenidi, L. and Antonio, R. A.},
	journal = {J. Fluid Mech.},
	pages = {395--420},
	title = {Effects of initial conditions in decaying turbulence generated by passive grids},
	volume = {585},
	doi={https://doi.org/10.1017/S0022112007006763},
	year = {2007}}

@article{zhao23_EJMBF,
	author = {Zhao, Y. and Yang, Y. and Li, M and Peng, Y.},
	journal = {Eur. J. Mech. B-Fluids},
	pages = {46--55},
	title = {Measurements of decaying grid turbulence with various initial conditions},
	volume = {102},
	doi={https://doi.org/10.1016/j.euromechflu.2023.07.005},
	year = {2023}}

@article{Nove_2020,
	author = {C. Nov{\'e}-Josserand and S. Perrard and A. Lozano-Duran and M. Benzaquen and M. Rabaud and F. Moisy},
	journal = {Phys. Rev. Fluids},
	pages = {124801},
	title = {Effect of a weak current on wind-generated waves in the wrinkle regime},
	volume = {5},
	year = {2020},
	doi={https://doi.org/10.1103/PhysRevFluids.5.124801}}

@article{Vellingiri2013,
	author = {Vellingiri, R. and Tseluiko, D. and Savva, N. and Kalliadasis, S.},
	journal = {Int. J. Multiphase Flow},
	pages = {93 - 104},
	title = {Dynamics of a liquid film sheared by a co-flowing turbulent gas},
	volume = {56},
	doi={https://doi.org/10.1016/j.ijmultiphaseflow.2013.05.011},
	year = {2013}}

@article{rabaud2013ship,
	author = {Rabaud, Marc and Moisy, Fr{\'e}d{\'e}ric},
	journal = {Phys. Rev. Lett.},
	number = {21},
	pages = {214503},
	publisher = {APS},
	title = {Ship wakes: Kelvin or {M}ach angle?},
	volume = {110},
	year = {2013},
	doi={https://doi.org/10.1103/PhysRevLett.110.214503}}

@article{Perrard2019,
	author = {Perrard, S. and Lozano-Dur{\'a}n, A. and Rabaud, M. and Benzaquen, M. and Moisy, F.},
	journal = {J. Fluid Mech.},
	pages = {1020--1054},
	publisher = {Cambridge University Press},
	title = {Turbulent windprint on a liquid surface},
	volume = {873},
	doi={https://doi.org/10.1017/jfm.2019.318 },
	year = {2019}}

@article{Paquier_2016,
	author = {Paquier, A. and Moisy, F. and Rabaud, M.},
	journal = {Phys. Rev. Fluids},
	pages = {083901},
	title = {Viscosity effects in wind wave generation},
	volume = {1},
	year = {2016},
	doi={https://doi.org/10.1103/PhysRevFluids.1.083901}}

@article{Paquier_2015,
	author = {Paquier, A. and Moisy, F. and Rabaud, M.},
	journal = {Phys. Fluids},
	pages = {122103},
	title = {Surface deformations and wave generation by wind blowing over a viscous liquid},
	volume = {27},
	year = {2015},
	doi={https://doi.org/10.1063/1.4936395}}

@article{Zhang_1995,
	author = {X. Zhang},
	journal = {J. Fluid Mech.},
	pages = {51-82},
	title = {Capillary--gravity and capillary waves generated in a wind wave tank: Observations and theories},
	volume = {289},
	doi={https://doi.org/10.1017/S0022112095001236},
	year = {1995}}

@article{Miles_1957,
	author = {Miles, J. W.},
	journal = {J. Fluid Mech.},
	pages = {185-204},
	title = {On the generation of surface waves by shear flows},
	volume = {3},
	doi={https://doi.org/10.1017/S0022112057000567},
	year = {1957}}

@article{Kahma_1988,
	author = {Kahma, K. and Donelan, M. A.},
	journal = {J. Fluid Mech.},
	pages = {339--364},
	publisher = {Cambridge Univ Press},
	title = {A laboratory study of the minimum wind speed for wind wave generation},
	volume = {192},
	doi={https://doi.org/10.1017/S0022112088001892},
	year = {1988}}

@article{aulnette2022kelvin,
	author = {Aulnette, M. and Zhang, J. and Rabaud, M. and Moisy, F.},
	journal = {Phys. Rev. Fluids},
	pages = {014003},
	publisher = {APS},
	title = {Kelvin-{H}elmholtz instability and formation of viscous solitons on highly viscous liquids},
	volume = {7},
	doi={https://doi.org/10.1103/PhysRevFluids.7.014003},
	year = {2022}}

@inproceedings{irps2016interaction,
	author = {Irps, T. and Kanjirakkad, V.},
	booktitle = {EPJ Web Conf.},
	pages = {02048},
	title = {On the interaction between turbulence grids and boundary layers},
	volume = {114},
	year = {2016},
	doi={10.1051/epjconf/201611402048},
}

@article{BANNER_1998,
	author = {Banner, M. L and Peirson, W. L.},
	journal = {J. Fluid Mech.},
	m3 = {10.1017/S0022112098001128},
	pages = {115--145},
	title = {Tangential stress beneath wind-driven air-water interfaces},
	ty = {JOUR},
	volume = {364},
	year = {1998},
	doi={https://doi.org/10.1017/S0022112098001128}}

@article{caulliez_2008,
	author = {Caulliez, G. and Makin, V. and Kudryavtsev, V.},
	journal = {J. Phys. Oceanogr.},
	pages = {2038--2055},
	title = {Drag of the water surface at very short fetches: Observations and modeling},
	volume = {38},
	year = {2008},
	doi={https://doi.org/10.1175/2008JPO3893.1}}

@article{fulgosi2003,
	author = {Fulgosi, M. and Lakehal, D. and Banerjee, S. and De Angelis, V.},
	journal = {J. Fluid Mech.},
	pages = {319--345},
	publisher = {Cambridge Univ Press},
	title = {Direct numerical simulation of turbulence in a sheared air--water flow with a deformable interface},
	volume = {482},
	year = {2003},
	doi={https://doi.org/10.1017/S0022112003004154}}

@article{Gastel1985phase,
	author = {van Gastel, K. and Janssen, P. and Komen, G. J.},
	journal = {J. Fluid Mech.},
	pages = {199--216},
	publisher = {Cambridge Univ Press},
	title = {On phase velocity and growth rate of wind-induced gravity-capillary waves},
	volume = {161},
	year = {1985},
	doi={https://doi.org/10.1017/S0022112085002889}}

@article{gottifredi_1970,
	author = {Gottifredi, J. and Jameson, G.},
	journal = {Proc. R. Soc. London Ser. A- Math. Phys. Eng. Sci.},
	number = {1538},
	pages = {373--397},
	publisher = {The Royal Society},
	title = {The growth of short waves on liquid surfaces under the action of a wind},
	volume = {319},
	year = {1970},
	doi={https://doi.org/10.1098/rspa.1970.0184}}

@article{grare2013growth,
	author = {Grare, L. and Peirson, W. and Branger, H. and Walker, J. and Giovanangeli, J-P. and Makin, V.},
	journal = {J. Fluid Mech.},
	pages = {5--50},
	publisher = {Cambridge Univ Press},
	title = {Growth and dissipation of wind-forced, deep-water waves},
	volume = {722},
	year = {2013},
	doi={https://doi.org/10.1017/jfm.2013.88}}

@article{hu2006wall,
	author = {Hu, Z. and Morfey, Ch. L. and Sandham, N. D.},
	journal = {AIAA J.},
	pages = {1541--1549},
	title = {Wall pressure and shear stress spectra from direct simulations of channel flow},
	volume = {44},
	year = {2006},
	doi={https://doi.org/10.2514/1.17638}}

@book{Janssen_2004,
	author = {P. Janssen},
	title = {The Interaction of Ocean Waves and Wind},
	publisher = {Cambridge University Press},
	year = {2004},
	}

@article{jimenez2008turbulent,
	author = {Jimenez, J. and Hoyas, S.},
	journal = {J. Fluid Mech.},
	pages = {215--236},
	publisher = {Cambridge Univ Press},
	title = {Turbulent fluctuations above the buffer layer of wall-bounded flows},
	volume = {611},
	year = {2008},
	doi={https://doi.org/10.1017/S0022112008002747}}

@article{Kawai1979generation,
	author = {Kawai, S.},
	journal = {J. Fluid Mech.},
	pages = {661--703},
	publisher = {Cambridge Univ Press},
	title = {Generation of initial wavelets by instability of a coupled shear flow and their evolution to wind waves},
	volume = {93},
	year = {1979},
	doi={https://doi.org/10.1017/S002211207900197X}}

@book{komen1996dynamics,
	author = {Komen, Gerbrand J and Cavaleri, Luigi and Donelan, Mark and Hasselmann, Klaus and Hasselmann, S and Janssen, PAEM},
	publisher = {Cambridge University Press},
	title = {Dynamics and Modelling of Ocean Waves},
	year = {1996}}

@book{Lamb,
	author = {Lamb, H.},
	publisher = {6th edition, Cambridge University Press},
	title = {Hydrodynamics},
	year = {1932}}

@article{Miles1959_part2,
	author = {Miles, J. W.},
	journal = {J. Fluid Mech.},
	pages = {568-582},
	title = {On the generation of surface waves by shear flows. {P}art 2.},
	volume = {6},
	year = {1959},
	doi={https://doi.org/10.1017/S0022112059000830}}

@article{Miles1993surface,
	author = {Miles, John},
	journal = {J. Fluid Mech.},
	pages = {427--441},
	publisher = {Cambridge Univ Press},
	title = {Surface-wave generation revisited},
	volume = {256},
	year = {1993},
	doi={https://doi.org/10.1017/S0022112093002836}}

@article{Miles1962generation,
	author = {Miles, John W},
	journal = {J. Fluid Mech.},
	number = {03},
	pages = {433--448},
	publisher = {Cambridge Univ Press},
	title = {On the generation of surface waves by shear flows. {P}art 4},
	volume = {13},
	year = {1962},
	doi={https://doi.org/10.1017/S0022112062000828}}

@article{Miles1967generation,
	author = {Miles, John W},
	journal = {J. Fluid Mech.},
	number = {01},
	pages = {163--175},
	publisher = {Cambridge Univ Press},
	title = {On the generation of surface waves by shear flows. {P}art 5},
	volume = {30},
	year = {1967},
	doi={https://doi.org/10.1017/S0022112067001351}}

@article{Moisy09,
	author = {F. Moisy and M. Rabaud and K. Salsac},
	journal = {Exp. Fluids},
	pages = {1021-1036},
	title = {A Synthetic {S}chlieren method for the measurement of the topography of a liquid interface},
	volume = {46},
	year = {2009},
	doi={https://doi.org/10.1007/s00348-008-0608-z}}

@article{PEIRSON:2008,
	author = {Peirson, W. L. and Garcia, A. W.},
	journal = {J. Fluid Mech.},
	pages = {243--274},
	title = {On the wind-induced growth of slow water waves of finite steepness},
	ty = {JOUR},
	volume = {608},
	year = {2008},
	doi={https://doi.org/10.1017/S002211200800205X}}

@article{phillips_1957,
	author = {Phillips, O. M.},
	journal = {J. Fluid Mech.},
	pages = {417--445},
	publisher = {Cambridge Univ Press},
	title = {On the generation of waves by turbulent wind},
	volume = {2},
	year = {1957},
	doi={https://doi.org/10.1017/S0022112057000233 }}

@article{plant1982relationship,
	author = {Plant, W. J.},
	journal = {J. Geophys. Res.-Oceans},
	pages = {1961--1967},
	publisher = {Wiley Online Library},
	title = {A relationship between wind stress and wave slope},
	volume = {87},
	year = {1982},
	doi={ https://doi.org/10.1029/JC087iC03p01961}}

@article{sullivan2010dynamics,
	author = {Sullivan, P. P. and McWilliams, J. C.},
	journal = {Annu. Rev. Fluid Mech},
	pages = {19--42},
	publisher = {Annual Reviews},
	title = {Dynamics of winds and currents coupled to surface waves},
	volume = {42},
	year = {2010},
	doi={https://doi.org/10.1146/annurev-fluid-121108-145541}}

@article{veron2001experiments,
	author = {Veron, F. and Melville, W. K.},
	journal = {J. Fluid Mech.},
	pages = {25--65},
	publisher = {Cambridge Univ Press},
	title = {Experiments on the stability and transition of wind-driven water surfaces},
	volume = {446},
	year = {2001},
	doi={https://doi.org/10.1017/S0022112001005638}}

\end{document}